\documentclass[pdfusetitle,aps,prd,nofootinbib,twocolumn,superscriptaddress,preprintnumbers]{revtex4-2}
\usepackage{amssymb}
\usepackage{amsmath,mathrsfs}
\usepackage{epsfig}
\usepackage{hyperref}
\usepackage{breakurl}
\usepackage{bm}
\usepackage{color}
\usepackage[dvipsnames]{xcolor}
\usepackage{tikz}
\usepackage[utf8]{inputenc}
\usepackage[normalem]{ulem}
\usetikzlibrary{decorations.markings}
\usepackage{tikz}
\usetikzlibrary{decorations.pathmorphing,arrows.meta}

\usepackage{bibunits}

\newcommand{\sug}[1]{#1}

\usepackage{multirow}

\usepackage{xcolor}
\usepackage{hyperref}
\hypersetup{
    colorlinks,
    linkcolor={red!40!black},
    citecolor={blue!40!black},
    urlcolor={blue!70!black}
}

\def\sectionskip{\vskip .2 cm}

\renewcommand{\imath}{\mathrm{i}}

\def\nn{\nonumber}
\def\Section#1{\noindent {\it #1}}

\newcommand{\order}{\mathcal{O}}
\newcommand{\Atree}{A^{(0)}}
\newcommand{\Aoneloop}{A^{(1)}}
\newcommand{\cL}{\mathcal{L}}
\newcommand{\cO}{\mathcal{O}}

\newcommand{\tr}[1]{\textrm{tr}\!\left( #1 \right)}

\newcommand{\Op}{\mathcal{O}}
\newcommand{\HD}{HD}
\newcommand{\Hb}{H\square}
\newcommand{\HB}{H^2B^2}
\newcommand{\HBd}{H^2\widetilde{B}B}
\newcommand{\HW}{H^2W^2}
\newcommand{\HWd}{H^2\widetilde{W}W}
\newcommand{\HG}{H^2G^2}
\newcommand{\HGd}{H^2\widetilde{G}G}
\newcommand{\HWB}{H^2WB}
\newcommand{\HWBd}{H^2\widetilde{W}B}
\newcommand{\W}{W^3}
\newcommand{\Wd}{\widetilde{W}W^2}

\newcommand{\lrb}{\left(}
\newcommand{\rrb}{\right)}
\newcommand{\lsb}{\left[}
\newcommand{\rsb}{\right]}

\begin{document}

\title{
\sug{Positivity in the Renormalization of Effective Field Theory}
}

\author{You-Peng Liao}
\affiliation{
Department of Physics and Center for Theoretical Physics,
National Taiwan University, Taipei 10617, Taiwan
}
\author{Jasper Roosmale Nepveu}
\affiliation{
Department of Physics and Center for Theoretical Physics,
National Taiwan University, Taipei 10617, Taiwan
}
\affiliation{Max Planck-IAS-NTU Center for Particle Physics, Cosmology and Geometry, Taipei 10617, Taiwan}
\author{Chia-Hsien Shen}
\affiliation{
Department of Physics and Center for Theoretical Physics,
National Taiwan University, Taipei 10617, Taiwan
}
\affiliation{Max Planck-IAS-NTU Center for Particle Physics, Cosmology and Geometry, Taipei 10617, Taiwan}
\affiliation{
Physics Division, National Center for Theoretical Sciences, Taipei 10617, Taiwan}
\affiliation{Department of Physics and Astronomy, Uppsala University, Box 516, 75120 Uppsala, Sweden}

\begin{abstract}
We show that the direction of renormalization in effective field theory is constrained by fundamental principles in the infrared---unitarity, analyticity, and Lorentz invariance. 
Our theorem, in the spirit of the $a$-theorem in conformal field theory, determines the sign of the one-loop running of couplings in the forward limit, when one inserts two operators whose mass dimensions are identical and even.
The theorem holds for a broad class of effective field theories with arbitrary ultraviolet completions.
The constraint directly applies to linear positivity bounds derived using tree-level amplitudes in the infrared, providing a criterion for whether renormalization effects can preserve the positivity bounds, or lead to their apparent violation.
We discuss the phenomenological implications of our theorem in
chiral perturbation theory and the Standard Model Effective Field Theory, where our theorem is particularly constraining for the running at dimension eight.
We provide several examples and show various extensions and applications even at dimension six.
\end{abstract}
   
\maketitle

\pdfbookmark[1]{Introduction}{intro}
\Section{Introduction.}
\label{sec:intro} 
Renormalization is a central aspect of field theory that governs how physical parameters evolve with energy scale.
The structures of the renormalization group (RG) equations---including, notably, the sign of the flow---carry crucial phenomenological implications~\cite{PhysRevLett.30.1343,PhysRevLett.30.1346}. 
Although the methods for renormalization are well established, it is often challenging to infer the behavior of the RG without performing explicit calculations.
Only a few exceptions are known, mainly in conformal field theory~\cite{Zamolodchikov:1986gt,Cardy:1988cwa}. 
A famous example is the four-dimensional $a$-theorem, which constrains
the $a$\nobreakdash-anomaly between the infrared (IR) and ultraviolet (UV) fixed points~\cite{Komargodski:2011vj,Komargodski:2011xv,Luty:2012ww},
\begin{align} 
a_{\textrm{IR}} - a_{\textrm{UV}} \le 0.
\label{eq:athm_CFT}
\end{align} 
This raises a natural question: do definite signs in RG exist for a broader set of parameters and field theories?

We address this question in the context of effective field theory (EFT).
The core concept of EFT is to replace UV physics at scale $\Lambda$ with a tower of local operators in the IR.
A classic example is chiral perturbation theory ($\chi$PT)~\sug{\cite{Weinberg:1966kf,Cronin:1967jq,Weinberg:1966fm,Weinberg:1968de,Gasser:1983yg,Gasser:1984gg}}
describing the strong interaction among low-energy \sug{mesons}.
In the search for physics beyond the Standard Model (SM), the Standard Model EFT (SMEFT) has become a prominent platform for systematically parameterizing new physics at the scale $\Lambda$ by extending the SM with higher-dimensional operators~\cite{Brivio_2019}.
While much of the focus is on dimension-6 operators at tree level, a variety of reasons motivate extensions to loop level~\cite{Panico:2018hal,Degrande:2020evl,Heinrich:2022gzl,DasBakshi:2024krs,Biekotter:2025nln} 
and to dimension eight~\cite{Azatov:2016sqh,Bi:2019phv,Remmen:2019cyz,Remmen:2020vts,Alioli:2020kez,Gu:2020ldn,Boughezal:2022nof,Grojean:2024tcw,Assi:2024zap,Liao:2024xel}. 
The dimension-6 RG is fully known~\cite{Jenkins:2013zja,Jenkins:2013wua,Alonso:2013hga,Alonso:2014zka}, as well as many sectors at dimension eight~\cite{AccettulliHuber:2021uoa,Chala:2021pll,Helset:2022pde,DasBakshi:2022mwk,DasBakshi:2023htx,Assi:2023zid,Boughezal:2024zqa,Bakshi:2024wzz,Assi:2025fsm}. \sug{Moreover, the community is making rapid progress in renormalization beyond one loop, e.g.~\cite{ deVries:2019nsu,Jenkins:2023bls,Born:2024mgz,DiNoi:2024ajj}.}
A major challenge in reaching higher orders is the rapid growth in complexity: even with one generation of fermions, the number of independent couplings increases from 84 at dimension six to 993 at dimension eight~\cite{Henning:2015alf}.
This complexity is compounded at loop level by the operator mixing due to RG.
This challenge motivates the search for generic signatures in the SMEFT that are agnostic to UV completions.

A key example of universal structures in EFT is provided by non-renormalization theorems~\cite{Alonso:2014rga,Elias-Miro:2014eia,Cheung:2015aba,Bern:2019wie,Jiang:2020rwz,Machado:2022ozb,Cao:2023adc}, which forbid large classes of operator mixing even in the presence of non-trivial Feynman diagrams. These seemingly miraculous cancellations become transparent in the on-shell approach~\cite{Caron-Huot:2016cwu}, see further developments in \sug{\cite{Bern:2020ikv,EliasMiro:2020tdv,Baratella:2020lzz,Baratella:2020dvw,Jiang:2020mhe,EliasMiro:2021jgu,AccettulliHuber:2021uoa,Baratella:2021guc,Baratella:2022nog,Bresciani:2024shu,Bresciani:2023jsu,Aebischer:2025zxg}}. Crucially, these results are only based on fundamental IR principles and hold for any UV completion.

Another prominent class of universal structures is the set of positivity bounds on EFT couplings~\cite{Pham:1985cr,Adams:2006sv}, which impose powerful constraints on the IR dynamics from basic assumptions in the UV.
In the SMEFT, these bounds could provide a promising avenue for setting theoretical 
priors in experimental analyses~\sug{\cite{Low:2009di,Zhang:2018shp,Zhang:2020jyn,Remmen:2019cyz,Remmen:2020vts,Remmen:2020uze,Remmen:2022orj,Zhang:2020jyn,Yamashita:2020gtt,Chen:2023bhu,Hong:2024fbl,Remmen:2024hry,Chakraborty:2024ciu}}.
The simplest bounds constrain the two-to-two forward amplitudes by truncating the EFT to tree level, writing $A=c_{4n}s^{2n-2}$ at dimension-$4n$, where $s$ is the Mandelstam variable. These ``tree-level positivity bounds'' take the form $c_{4n} \ge 0$.
Importantly, these tree-level bounds are susceptible to loop corrections from light particles, first discussed in $\chi$PT~\cite{Pennington:1994kc,Ananthanarayan:1994hf,Dita:1998mh,Distler:2006if,Manohar:2008tc} and revived recently by \cite{Arkani-Hamed:2020blm,Bellazzini:2020cot,Bellazzini:2021oaj,Beadle:2024hqg,Beadle:2025cdx,Chang:2025cxc,Desai:2025alt,Chala:2021wpj,Li:2022aby,Ye:2024rzr},
which are relevant in the SMEFT as well.
For phenomenological applications, it is thus essential to account for loop effects before interpreting tree-level positivity bounds as robust theoretical constraints.

In this paper, we analyze the interplay between RG and tree-level positivity bounds for arbitrary spins.
We show that there is an analogue of the $a$-theorem in EFT: the sign of one-loop RG for $c_{4n}$, arising from the insertion of two dimension-$(2n+2)$ operators with $n\ge 2$, is fixed, \sug{a property we dub ``RG positivity'':
\begin{align}
    \left(c_{4n,\textrm{IR}} - c_{4n,\textrm{UV}}\right)\big\vert_{(\textrm{dim-$(2n+2)$})^2} 
    \ge 0\,.
    \label{eq:atheorem_simp}
\end{align} }
The precise statement will be given later around \eqref{eq:athm_precise}.
Although our theorem applies in the perturbative regime,
the proof does not use dispersion relations and relies only on fundamental principles in the IR: unitarity, analyticity, and Lorentz invariance.
It therefore applies to a wide class of theories, including $\chi$PT, the SMEFT, 
and EFTs for quantum gravity.

\begin{figure}[t]
\begin{center}
\includegraphics[trim={0 27.0cm 48cm 0},clip,scale=0.388]{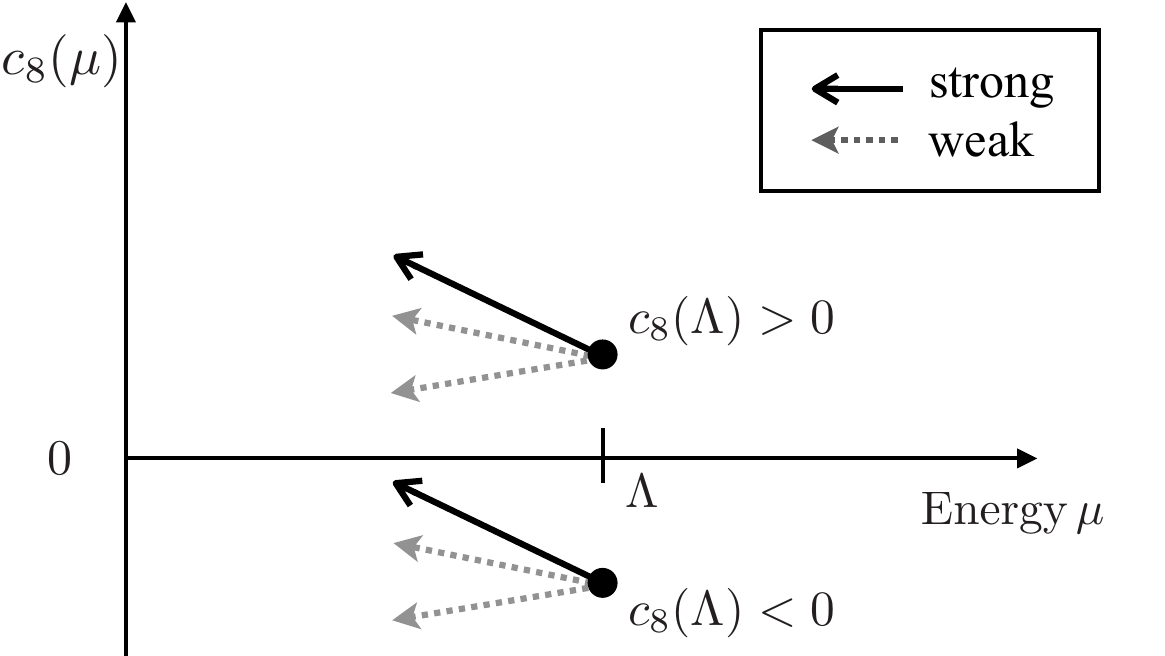}
\end{center}
\vspace{-0.4cm}
\caption{\label{fig:RG_plot} 
A schematic plot for the RG of a dimension-8 coupling $c_8$ subject to 
our theorem in Eq.~\eqref{eq:atheorem_simp}.
The solid arrows indicate the running of $c_8(\mu)$ 
in a ``strongly coupled'' EFT, 
as per
the criterion~\eqref{eq:strong_condition}, 
so that
the RG direction is controlled by~\eqref{eq:atheorem_simp}.
The grey arrows show weakly-coupled EFTs 
in which the running can go either way.
We consider two 
values of $c_8$ at the matching scale $\Lambda$; see examples with $c_8(\Lambda) \le 0$ in~\cite{Chala:2021wpj}.
}
\end{figure}

One can view 
our theorem~\eqref{eq:atheorem_simp} 
as a classification of the RG sectors that 
preserve the tree-level positivity of $c_{4n}$. 
For the SMEFT, this mainly applies to dimension-$8$ RG from inserting two dimension-$6$ operators, but not to the contribution from the mixing of 
dimension-8 and SM interactions.
This is consistent with the violation of tree-level positivity found in~\cite{Chala:2021wpj}; see also the examples in the supplemental material~\cite{Supp}\nocite{Catani:1996vz}.
Later we give a concrete criterion~\eqref{eq:strong_condition} for the preservation or violation of tree-level positivity by RG. See Fig.~\ref{fig:RG_plot} for a schematic plot.

This study is deeply influenced by the seminal results 
in~\cite{Chala:2021pll,Chala:2021wpj}, which have first observed non-trivial examples of 
 positivity in the RG
at dimension eight in the SMEFT.
The authors explained the structure as a consequence of positivity bounds from the UV.
See also~\cite{DasBakshi:2022mwk,DasBakshi:2023htx,Li:2022aby,Chala:2023jyx,Chala:2023xjy,Ye:2024rzr} for further discussions.
Our approach differs from the literature in several ways. The most important difference is that the proof of \eqref{eq:atheorem_simp} is solely based on fundamental principles in the IR, without any assumption on UV completions.
As a result, our theorem applies very generally, including EFTs with gravitons and operators beyond dimension eight.
Our derivation also yields extensions at dimension six, which is typically free from positivity bounds.
A more detailed comparison with previous work will be presented later.

\begin{figure}[t]
\begin{center}
\includegraphics[trim={0 16cm 37cm 0},clip,scale=0.20]{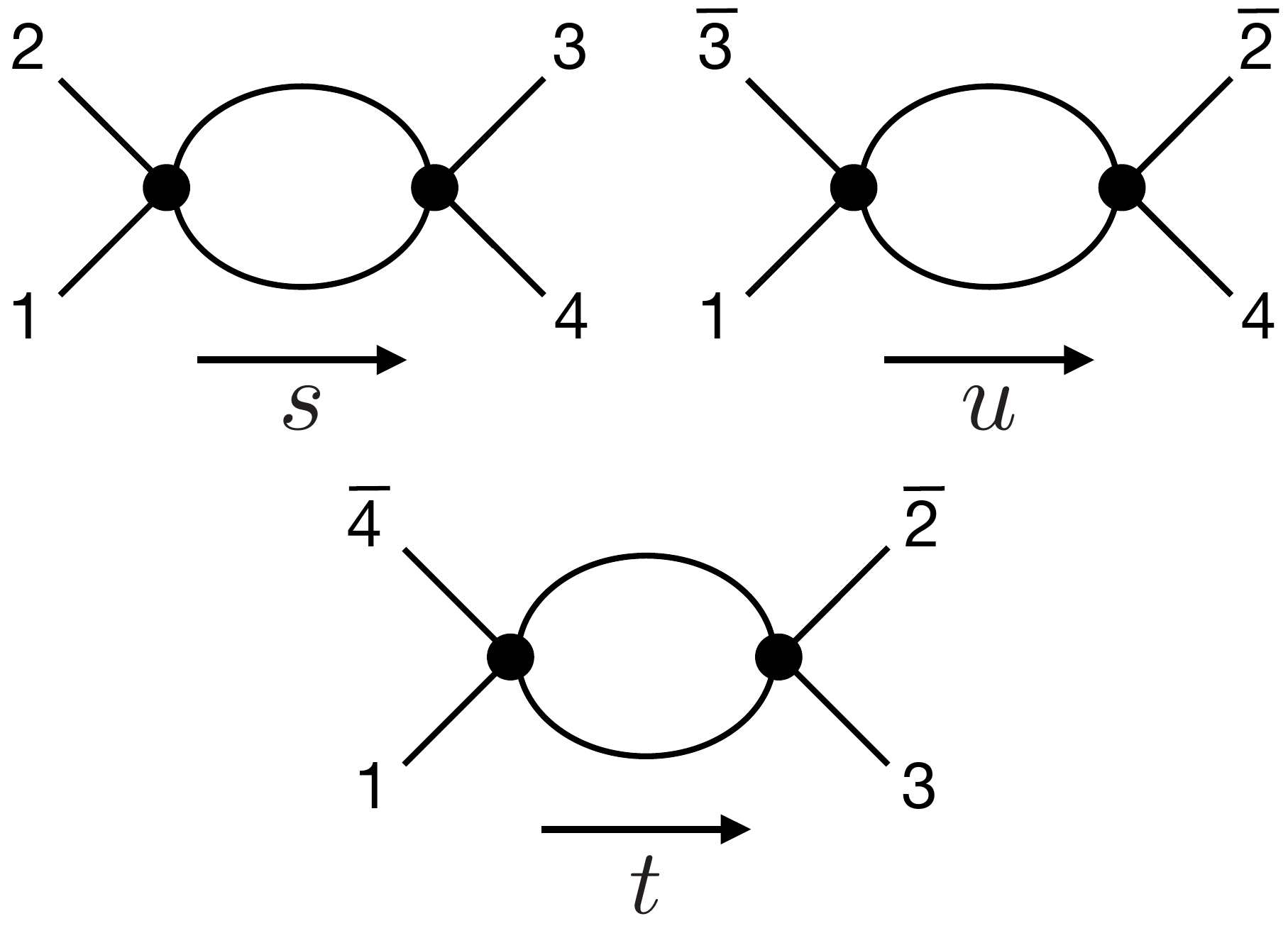}
\end{center}
\vspace{-0.5cm}
\caption{\label{fig:unitarity} 
The unitarity cuts of the $s$ and $u$ channels (top) and the $t$ channel (bottom). All the lines are on shell. 
We label the incoming states (1 and 2) and outgoing states (3 and 4) by their particle number, and the bar denotes an anti-particle.
In the FL, we take the states $|3\rangle=|2\rangle$ and $|4\rangle=|1\rangle$. 
In this case, the $s$ and $u$ channels have the same initial and final states, but the $t$ channel does not.
}
\end{figure}

\sectionskip
\pdfbookmark[1]{Proof of Equation (2)}{proof}
\Section{Proof of \sug{RG Positivity}.} 
\label{sec:proof}
We extract the RG in dimensional regularization from the on-shell formalism~\cite{Caron-Huot:2016cwu} 
to leverage the connection with fundamental principles.
To avoid complications from IR divergences and tadpole diagrams, we restrict to EFTs with massless particles without cubic interactions.
This still allows a large range of interactions, such as \sug{pions in $\chi$PT in the chiral limit}. For the SMEFT, the only higher-dimensional operators we neglect are the $F^3$ operators at dimension six.
We use $\phi,\psi$, and $F$ for scalars, fermions, and gauge field strength tensors, and we specify operators only by their field content, while omitting details of Lorentz and other indices.

Let us quickly review the on-shell methods.
We focus on the RG of a four-point operator renormalized by two four-point operators. 
The extraction of RG follows from the fact that the full amplitudes are independent of the renormalization scale $\mu$,
\begin{align}
    \frac{d}{d\ln\mu}A^{\rm full} =0 \,. 
\end{align}
To one-loop level, $A^{\rm full}=\Atree+\Aoneloop$ where the tree amplitude $\Atree$ is written in terms of the renormalized couplings $c_{i}(\mu)$. 
This implies $\tfrac{d\Atree}{d\ln\mu}= -\tfrac{d\Aoneloop}{d\ln\mu}$, where the leading $\mu$-dependence at one loop comes in the form of $\ln(s_i/\mu^2)$ where $s_i=s,t,u$ are the 
Mandelstam variables. 
The scale dependence of the couplings in $A^{(1)}$ is a two-loop effect.
Crucially, the coefficients of these logarithms can be computed from 
the branch cut of each $s_i$,%
\footnote{
    Instead of using form factors~\cite{Caron-Huot:2016cwu}, we use the kinematics $s_i >0$ to compute the discontinuity of each $\ln s_i$, then analytic continue to the same kinematics to combine all channels.
    This analytic continuation is trivial since the RG is given by the coefficient of UV divergence which is local, i.e., a polynomial in $s,t,u$.
    We emphasize that, \emph{a priori}, the branch cuts of all channels should be included in the calculation. For instance, this is needed to yield the correct RG in a $\lambda \phi^4$ theory.
}
\begin{align}
    \frac{d}{d\ln\mu} \Atree
	=& \,\frac{i}{\pi} \sum_{i=s,t,u} \,\textrm{Disc}_i\, \Aoneloop \label{eq:onshellRG_master} \\
    &\hspace{-6mm}= -\frac{1}{\pi} \sum_{i=s,t,u} \int_{\varphi''_i}\, 
    \langle \varphi_{i}|\hat{A}^{(0)}|\varphi''_i\rangle 
    \langle \overline{\varphi}''_i| \hat{A}^{(0)} |\varphi'_i \rangle \,, \nn
\end{align}
where we use the optical theorem in the last line to recast the discontinuity into unitarity cuts in terms of tree amplitudes integrated over the phase space $\int_{\varphi''_i}$ of the intermediate state $|\varphi''_i\rangle$.
We define $\Atree_{L,i} \equiv \langle \varphi_{i}|\hat{A}^{(0)}|\varphi''_i\rangle$ and $\Atree_{R,i}{}^\dagger \equiv \langle \overline{\varphi}''_i| \hat{A}^{(0)} |\varphi'_i \rangle $.
In our case, $\Atree_{L,i}$, $A^{(0)}_{R,i}$, and $\Atree$ are all four-point local amplitudes,
see the diagrams in Fig.~\ref{fig:unitarity}.
This formula succinctly derives the one-loop RG from tree amplitudes in the absence of IR divergences. Crucially, the RG derivation does not use dispersion relations and is therefore agnostic about UV completions.

The on-shell RG formula \eqref{eq:onshellRG_master} holds for generic kinematics. 
However, our main result is that additional structure can be exposed by considering special limits. In particular,
we take the forward limit (FL) where the incoming particles 1 and 2 have the same momenta and quantum numbers as the outgoing particles 4 and 3, i.e., their states satisfy $|3\rangle=|2\rangle$ and $|4\rangle=|1\rangle$.
This implies $t=(p_1-p_4)^2\rightarrow 0$, see Fig.~\ref{fig:unitarity}.
In the FL, the tree amplitude behaves as 
\sug{\begin{align}
     \Atree = \sum_{m=0}  
     c_{2m+4}
     s^{m},
     \label{eq:A_FL}
\end{align}}
where the coefficient 
\sug{$c_{2m+4}$} depends on the EFT parameters.
By defining \sug{$c_{2m+4}$} through the on-shell amplitudes, we have (partially) fixed the choice of operator basis.
The formula \eqref{eq:onshellRG_master} then computes the RG of \sug{$c_{2m+4}$},
thereby providing 
direct insight into the RG corrections to tree-level positivity bounds, \sug{$c_{2m+4} \ge 0$}.
In the following, 
it is a key property of the $s$ and $u$ channels that 
they have the same initial and final states 
in the FL, 
$|\varphi_i\rangle=|\varphi_i'\rangle$, 
while this is not true in the $t$ channel. Let us consider these cases separately.

For the $s$ and $u$ channels, the total integrand in \eqref{eq:onshellRG_master} is positive definite due to unitarity in their respective physical regimes (when $s$ or $u$ is positive).
However, the RG for $c_{2m+4}$ requires us to split the right-hand side of \eqref{eq:onshellRG_master} into a power series in $s$ which may break positivity of individual couplings.
The direction of the RG running for $c_{2m+4}$ then hinges on the truncation of \eqref{eq:onshellRG_master} to  order $s^m$, as well as the combination between the $s$ and $u$ channels.
First, we impose that both $\hat{A}^{(0)}$ come from operators of the same mass dimension \sug{$(m+4)$},
such that each integrand is always non-negative:
$\int_{\varphi''_i} \Atree_{L,i} A^{(0)\dagger}_{R,i} = d_{i,m}\,s_i^m,$ for a non-negative coefficient $d_{i,m}$ and $s_i=s$ or $u$.
Second, since $s=-u$, the sum over the two channels is 
manifestly
sign-definite only when $m$ is even.\footnote{
The same constructive interference also appears in 
the derivation of tree-level positivity bounds.
}
To satisfy both criteria, we find that the contribution from these channels has a definite sign when we insert a pair of operators whose mass dimensions are even and equal.
For instance, we have control on the RG of dimension-8 operators from a pair of dimension-6 operators but not from the interference between dimension-4 and dimension-8 operators, nor the RG of dimension-6 operators from a pair of dimension-5 operators.

For the $t$ channel, the integrand in \eqref{eq:onshellRG_master} is not necessarily positive in the FL since $|\varphi_i\rangle \neq |\varphi'_i\rangle$.
We instead prove that $\textrm{Disc}_t\, \Aoneloop =0$ in the FL when inserting any higher-dimensional operator.\footnote{
Setting $t=0$ directly leads to vanishing scaleless bubbles. However, for the extraction of the RG,  
we need to show that the \emph{coefficient} of the scaleless bubble, $\textrm{Disc}_t\, \Aoneloop$, vanishes instead.}
Consider the two internal particles to have momenta $\ell_j$ and helicity $h_j$.
Recall that a local amplitude containing a particle with helicity $-|h|$ or $|h|$ must be proportional to $|p\rangle^{2h}$ or $|p]^{2h}$ in the spinor-helicity variables. Any extra derivative on the particle adds another factor of $|p\rangle [p|$. Applying this to the local amplitudes $\Atree_{L,i}$ and $A^{(0)\dagger}_{R,i}$ in \eqref{eq:onshellRG_master} yields
\begin{align}
    \textrm{Disc}_t \Aoneloop
    \propto & \int_{\varphi''_t}\, \left(|\ell_1 \rangle [\ell_1| \right)^{2|h_1|+n_1}
    \left(|\ell_2 \rangle [\ell_2| \right)^{2|h_2|+n_2} \nn \\
    \propto & \, (p_1-p_4)^{2|h_1|+2|h_2|+n_1+n_2},
    \label{eq:t_channel}
\end{align}
where $n_{i}$ are the number of derivatives on $\ell_{i}$ and we only keep track of the $\ell_{i}$ dependence but do not assume any details on index contractions.
In the second line, we crucially use Lorentz invariance to replace the internal  $\ell_{1,2}$ vectors by the external momentum $p_1-p_4$ after integration. It is important that we only have a bubble diagram so that the phase-space integration involves a single scale.
This shows that $\textrm{Disc}_t \Aoneloop$ can be non-zero only when both internal lines are scalar particles without any extra derivative.
But in such a case, all external momenta or spinors must contract with the ones from particles on the same side of the unitarity cut which vanish in the FL, due to $p_i^2=0$ and $\langle i i \rangle = [ii]=0$.
This singles out $\phi^4$-theory as the only case where $\textrm{Disc}_t \Aoneloop \neq 0$ in the forward limit.%
    \footnote{One can also see this from the partial wave perspective~\cite{Baratella:2021guc}, noting that in the $t$-channel partial-wave decomposition, only the $J=(n-4)/2$ coefficient survives the FL of a dimension-$n$ operator.}
In other words, the $t$ channel does not contribute to the RG of any higher-dimensional operator which survives the FL,
given that $\Atree_{L,i}$ and $\Atree_{R,i}$ are local. Notably, this includes dimension-6 operators.

Let us combine the analysis for all channels in \eqref{eq:onshellRG_master}.
The $s$ and $u$ channels for the RG of $c_{4n}$ induced by two dimension-$(2n+2)$ operators have definite sign, and the $t$-channel contribution vanishes for $n \ge 2$.
We have thus proven \sug{the RG positivity:
\begin{align}
    \frac{d\,c_{4n}}{d\ln \mu} \bigg\vert_{(\textrm{dim-}(2n+2))^2} \le 0\,.
    \label{eq:athm_precise}
\end{align} }
While examples of this theorem are known in the scalar case~\cite{Arkani-Hamed:2020blm,Bellazzini:2020cot,Bellazzini:2021oaj} by explicit calculations
and at dimension eight in the SMEFT~\cite{Chala:2021pll,Chala:2021wpj}, our proof holds for arbitrary spins and mass dimension.
This theorem leads to the solid lines in Fig.~\ref{fig:RG_plot} for the running of $c_{4n}$.
The vanishing of the $t$-channel discontinuity for higher-dimensional operators also leads to a corollary for a new type of non-renormalization theorems which we will discuss later.

As we have emphasized, our proof relies only on unitarity, analyticity, and Lorentz invariance in the \emph{IR} EFT regime, while being agnostic on the UV completion. This is in contrast to the argument for the signs and zeros in the RG at the dimension-8 level in~\cite{Chala:2021wpj}, which uses positivity assumptions in the \emph{UV}, 
together with a
classification of operators in weakly-coupled UV completions.
The \sug{RG positivity} dictated by~\eqref{eq:athm_precise} thus explains and generalizes the results of \cite{Chala:2021wpj} from a pure IR perspective.

\sectionskip
\pdfbookmark[1]{Phenomenological Implications}{pheno}
\Section{Phenomenological Implications.}\label{sec:app}
The generality of the above derivation implies that   our result
applies to a wide range of EFTs. Here, we discuss the phenomenological implications for $\chi$PT and the SMEFT.

We use the $\chi$PT Lagrangian at $\order(p^2)$ given by 
$\cL_2= f^2\,\tr{\partial_\mu U^\dagger \partial^\mu U }/4$,  where $U=\exp{(2iT^a \pi^a/f)}$, with 
$\tr{T^a T^b}=\delta^{ab}/2$. 
The $\order(p^4)$ Lagrangian with $SU(2)_V$ symmetry is then
$\cL_4= c_1 \,\tr{\partial_\mu U^\dagger \partial^\mu U }^2
+c_2 \, \tr{\partial_\mu U^\dagger \partial_\nu U }
\tr{\partial^\nu U^\dagger \partial^\mu U }$~\cite{Gasser:1983yg}, which yields the tree-level forward amplitude
\sug{\begin{equation}
    A_{ij\to ji}
    =\begin{cases}
    16\,s^2\,(c_1+c_2)    &\text{if }i=j\\
    8\,s^2\,c_2    &\text{if }i\neq j,
    \end{cases}
\end{equation}
where
 $i,j$ label the flavor of pions in the $s$ channel.} 
The \sug{RG positivity}~\eqref{eq:athm_precise}
requires $\frac{d\,c_2}{d\ln \mu} \le 0$ and $\frac{d(c_1+c_2)}{d\ln \mu} \le 0$ when considering all pion species. Indeed, this agrees with the explicit RG calculation:
\begin{align}
    \frac{d\,(c_1+c_2)}{d\ln \mu}  &= -
    \frac{1}{16\pi^2}
    \frac{1}{4}\,,
    &\frac{d\, c_2}{d\ln \mu} &= -
    \frac{1}{16\pi^2}
    \frac{1}{6}\,.
    \label{eq:chiPT}
\end{align}
Moreover, the loop corrections induced by the $O(p^2)$ Lagrangian are naturally comparable to the tree-level $O(p^4)$ terms~\cite{Manohar:1983md}, leading to $\mathcal{O}(c_i) \sim 0.01$.
Crucially, the scale dependence would exhibit the same qualitative behavior in the presence of electromagnetic interactions between charged pions, whose effects on \eqref{eq:chiPT} are negligible because 
$\mathcal{O}(\tfrac{\alpha_\textsc{em}}{4\pi})\times \mathcal{O}(c_i) \lesssim 10^{-5}$.
Therefore, our theorem~\eqref{eq:athm_precise}
phenomenologically captures the direction of the RG flow even though it does not determine the sign of all terms.

Let us turn to the SMEFT. The dimension-8 RG in the SMEFT in the unbroken phase takes the form 
\begin{equation}\label{eq:6vs8}
    \frac{d\,c_{8,i}}{d\ln\mu}  = \gamma_{ij}\,c_{8,j} + \gamma'_{ijk}\,c_{6,j}\,c_{6,k}\,,
\end{equation}
where $c_{n,i}$ denotes a coupling associated with a dimension-$n$ operator $\cO_{n,i}$.
The anomalous dimension $\gamma$ and $\gamma'$ may depend on the SM couplings.
The sign of the $\gamma'$ term
is determined by \sug{RG positivity}
if $c_{8,i}$ can be probed in the FL.%
    \footnote{Note that further definite signs and zeros were identified in some entries of $\gamma$ in~\cite{Chala:2023jyx,Chala:2023xjy}.
}
For tree-level scattering in the SMEFT, the appearance of $c_{8,i}$ in the FL has been systematically studied in~\cite{Remmen:2019cyz}, where the combination $c_{H^4D^4}^{(1)}+c^{(2)}_{H^4 D^4}$ appears as one of them.%
    \footnote{
    We use $\cL_8=\sum_i c^{(i)}_{H^4D^4} \cO^{(i)}_{H^4D^4}/\Lambda^4$, where $\cO^{(i)}_{H^4D^4}\equiv \cO^{(i)}_{H^4}$ in the basis of~\cite{Murphy:2020rsh}.
    }
Thus, the $\gamma'$ for this combination is indeed negative:
\begin{align}
    &16\pi^2\frac{d \left(c_{H^4D^4}^{(1)}+c^{(2)}_{H^4 D^4}\right)}{d\ln \mu}  \\
    =&
-16\left( c_{ H^2WB}^2
+c_{ H^2\tilde{W}B}^2\right)
-12\,g_2^2
\left( c_{\W}^2 + c_{\Wd}^2\right)
+\dots,\nn
\end{align}
where we only show the gauge boson loops.%
    \footnote{
    We use $\cL_6=\sum_i c_{i} \cO_{i}/\Lambda^2$, where $\cO_i$ is defined in the basis of~\cite{Grzadkowski:2010es} 
    as
    $\cO_{H^2 XB}\equiv \cO_{\varphi XB}$, 
    $\cO_{W^3}\equiv \cO_{W}$, and
    $\cO_{\widetilde{W}W^2}\equiv \cO_{\widetilde{W}}$.
    }
Remarkably, double insertions of the $F^3$ operators also mix into $H^4D^4$ with a negative sign, even though we assumed the absence of cubic vertices in our proof. In the present case, we only have bubble diagrams,
\begin{equation}
 \includegraphics[page=6,trim={0 29.6cm 49cm 0cm},clip,scale=0.2]{Diagrams} 
 \raisebox{9.5mm}{\hspace{-3mm}+ $t$-channel\,,}
 \vspace{-2mm}
\end{equation}
so the $t$-channel cut still depends on a single scale and must vanish by the same counting as in \eqref{eq:t_channel}.
We thus conclude that the \sug{RG positivity} constrains the sign of the full RG of $H^4D^4$ operators induced by two dimension-6 insertions.
We provide this RG in the supplemental material~\cite{Supp}.%
\footnote{\sug{In the supplemental material~\cite{Supp}, we also discuss an example with three-point interactions, to which our derivation may not apply. One can see that the optical theorem does not imply a definite sign of the UV divergence. Thus, the systematic extension of our theorem to include three-point interactions requires further investigation that is beyond the scope of this paper.}}
Together with the results of~\cite{Chala:2021pll,
Helset:2022pde}, this completes the RG of the $H^4D^4$ operators from double insertions of dimension-6 operators.

The relative importance of the two terms in \eqref{eq:6vs8} depends on the UV completion. 
The contribution from a pair dimension-6 operators is dominant whenever the EFT is more strongly coupled than the SM, 
\begin{align}
    \order\!\left(\widehat{c}_{6}\right)^2 \gg \order\!\left(\widehat{c}_{8}\right) \order\!\left(\widehat{c}_{4}\right),
    \label{eq:strong_condition}
\end{align}
where $\order(\widehat{c}_{n})$ is the order of a dimension-$n$ coefficient normalized according to naive dimensional analysis~\cite{Manohar:1983md,Jenkins:2013sda}.
$\order(\widehat{c}_{4})$ is known to be small since the SM is weakly coupled above the electroweak scale.
If \eqref{eq:strong_condition} holds, 
which is natural in a range of UV completions~\cite{Giudice:2007fh,Biekotter:2014gup,Liu:2016idz,Contino:2016jqw,Stefanek:2024kds},
the dimension-8 couplings increase by running to the IR.

Conversely, if the heavy particles are (very) weakly coupled to the SM, the Wilson coefficients are naturally also small, in which case 
the \sug{RG positivity} theorem no longer fixes
the overall sign of the running at dimension-8 level.
This may even lead to violations of the tree-level positivity bounds in the IR~\cite{Chala:2021wpj}.
See Fig.~\ref{fig:RG_plot} for a schematic plot of the RG flow in various UV scenarios.

We emphasize that violations of tree-level positivity bounds on dimension-8 couplings are consistent with the constraints from dispersion relations. 
As explained in \cite{Arkani-Hamed:2020blm}, even though $c_{4n}$ in \eqref{eq:A_FL} is positive in a mass-deformed theory below the mass gap, such a deformation is not small at low energies and may lead to different behavior in the deep IR compared to massless theories.
Alternatively, an arc with a smaller radius in the picture of~\cite{Bellazzini:2020cot} is more positive, since the discontinuity is positive definite. 
However, the arc may be dominated by the loops of dimension-4 couplings~\cite{Chala:2021wpj,Li:2022aby,Chala:2023jyx,Ye:2024rzr}, such that the positivity of the arc does not constrain the sign of $c_{4n}$.
We exemplify the violation of tree-level positivity bounds in the supplemental material~\cite{Supp}, including a comparison to the dispersion relations.

\begin{figure}
\begin{center}
\begin{equation*}
    \vcenter{\hbox{\includegraphics[page=2,
    trim={0 27.0cm 50cm 0.1cm},clip,scale=0.2]{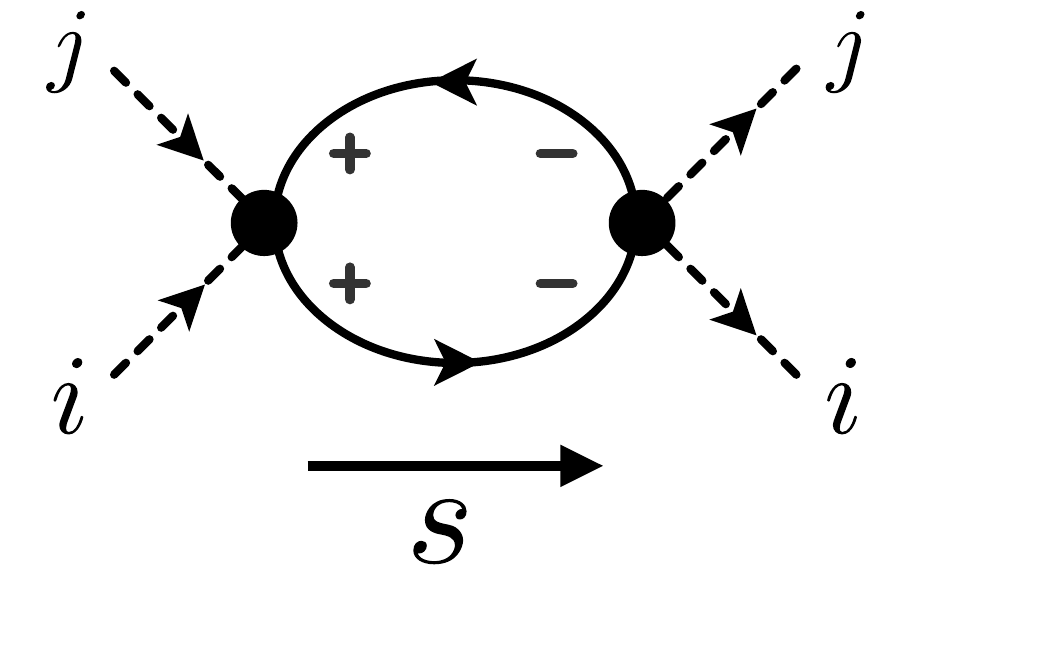}}}
    \hspace{-4mm}
    \raisebox{3.5mm}{$\;\sim\;
    \left(d_1\,\delta_{ij} + d_2\right)\,s\,.$}
    \vspace{-7mm}
\end{equation*}
\end{center}
\caption{\label{Fig:5squared} 
The only non-zero channel in the RG formula \eqref{eq:onshellRG_master} for the mixing of a pair of dimension-5 operators into the $H^4D^2$ operators in the SMEFT. The dashed and solid lines correspond to Higgs boson and lepton doublets, respectively. The $i,j$ denote $SU(2)$ indices of the external Higgs doublets. Helicities are indicated in the all outgoing convention. It follows from the \sug{RG positivity} that the constants satisfy $d_2 \geq 0$ and $d_1+d_2 \geq 0$, as verified by Eq.~\eqref{11}.
}
\end{figure}

\sectionskip
\pdfbookmark[1]{Further Applications}{applications}
\Section{Further Applications.}\label{sec:app}
The \sug{RG positivity} does not generally apply to the RG of dimension-6 operators, due to potential cancellations between the $s$ and $u$ channels. 
However, whenever one (or both) of these channels is absent, definite signs or zeros are still implied. For instance in the SMEFT, the RG of the $H^4D^2$ operators (using the basis of \cite{Grzadkowski:2010es}) induced by double insertions of the Weinberg operators $\psi^2H^2$~\cite{Weinberg:1979sa}  
only have the $s$ channel shown in Fig.~\ref{Fig:5squared}. 
Thus, the $\psi^2H^2$ contributions to the RG of $H^4D^2$ operators are sign-definite,%
    \footnote{
    In the FL, the dimension-6 four-Higgs amplitude is 
    $A_{ij\to ij} = (c_{HD} +2\,\delta_{ij} c_{H\square})\, s$, where $i,j$ are the $SU(2)$ indices of the Higgs doublets. This implies
    $\dot c_{HD}<0$ and $ \dot c_{HD} + 2\, \dot c_{H\square} <0$ from Weinberg operators.
    } 
as confirmed by the explicit results in~\cite{Broncano:2004tz,Davidson:2018zuo}
\begin{align}
   \frac{d\,c_{HD} }{d\ln\mu}  &=
   -\frac{4}{16\pi^2} \, \tr{c_5 c_5^*}
   \,, \nn\\
   \frac{d\,(c_{HD} + 2\,c_{H\square})}{d\ln\mu}  &= 
   -\frac{8}{16\pi^2} \, \tr{c_5 c_5^*}, \label{11}
\end{align}
where the trace sums over the flavor space. We also analyze the full mixing of the Weinberg operators into other dimension-6 operators in the supplemental material~\cite{Supp}.

A corollary of \eqref{eq:t_channel} is the non-renormalization of a higher-dimensional operator that can be probed in the FL if the $t$-channel cut is the only available one.
This predicts, for example, the non-renormalization of dimension-8 operators of the form $H^2 F^2 D^2$ from inserting $H^4D^2$ and $H^2F^2$ types of operators.
Even at dimension six, this explains the zero contribution from the $\lambda H^4$ potential to the diagonal anomalous dimension of the $\bar\psi \psi H^2D$ operators~\cite{Jenkins:2013zja}.
In contrast, the $H^2F^2$ operators vanish in the FL so the corollary does not apply.

An alternative perspective on the non-renormalization of 
$\bar\psi \psi H^2 D \to \bar\psi \psi H^2 D $ follows from the angular momentum selection rules identified in \cite{Jiang:2020rwz,Baratella:2021guc}. Namely, in the $\bar\psi \psi \to HH$ channel, these operators generate amplitudes with angular momentum (spin) $J=1$, while $\lambda H^4$ has spin zero and is thus orthogonal. 
The same argument applies to the non-renormalization of dimension-8 operators that survive in the FL by those that vanish in this limit (again, if only the $t$ channel exists)~\cite{Chala:2023jyx,Chala:2023xjy}.
This hinges on the fact that only the former class of operators has a non-zero $J=(n-4)/2$ $t$-channel partial wave.%
    \footnote{
    More intricate are the definite signs in the RG mixing among a certain class of dimension-8 operators with non-zero FL~\cite{Chala:2023jyx,Chala:2023xjy}. 
    We postpone a detailed study of these signs to future work.
    }

\sectionskip
\pdfbookmark[1]{Conclusion}{conclusion}
\Section{Conclusion.}\label{sec:outlook}
We systematically analyzed the direction of the one-loop RG flow of EFT couplings that are subject to tree-level positivity bounds.
The \sug{RG positivity}, derived solely from basic principles in the IR,  singles out the contributions that preserve the positivity at the matching scale under RG flow, but does not exclude potential violation from other sectors~\cite{Chala:2021wpj}.
If the violation of tree-level positivity bounds in the SMEFT are found experimentally, our results select weakly-coupled UV models as one of the possible scenarios.
We also discuss applications to the running of dimension-6 couplings and we derive new non-renormalization theorems.

Our work opens several promising avenues for future studies. 
A natural extension is to include cubic interactions in both gauge and gravitational theories. These contributions introduce additional complexity due to infrared divergences, though the positivity structure may still persist~\cite{Baratella:2021guc}, potentially offering sharper insights into the Weak Gravity Conjecture in the deep IR~\cite{Arkani-Hamed:2021ajd}.

It is also of interest to generalize these constraints to higher-point scattering, higher-loop corrections and massive particles, as well as to the two-point spectral density. 
\sug{For constraints on the RG of higher-point operators,  it is crucial to judiciously generalize the forward limit, 
see e.g.~\cite{Chandrasekaran:2018qmx,Cheung:2025nhw}.}
Furthermore, as we have focused on linear positivity bounds, another natural extension would be to apply similar methods to non-linear bounds~\cite{Arkani-Hamed:2020blm,Bellazzini:2020cot,Bellazzini:2021oaj}.
Furthermore, while we have illuminated
important constraints on the RG flow of couplings, it would be valuable to study the behavior of UV matching using on-shell methods~\cite{DeAngelis:2023bmd}.
Our theorem may also lead to interesting interplay when combined with the geometric renormalization~\cite{Alonso:2015fsp,Alonso:2016oah,Helset:2022tlf,Helset:2022pde,Assi:2023zid,Li:2024ciy,Aigner:2025xyt}.

Given the deep connection between RG flow and the counting of degrees of freedom, it would be intriguing to consider whether the \sug{RG positivity theorem}
can be reformulated from quantum information perspective~\cite{Aoude:2024xpx,Low:2024mrk,Low:2024hvn,DuasoPueyo:2024usw}.
Even decades after its development, we still have much to learn from the structures and implications of renormalization.

\Section{Acknowledgments.}                      
The authors thank C.-H.~Chang, M.~Chala, C.~Cheung, S.~De Angelis, G.~Durieux, J.~Gu, P.-S.~Chiu,
G.~Guedes, Y.-T.~Huang, A.~Manohar, J.~Parra-Martinez, G.~Remmen and N.~Rodd for stimulating discussions and helpful comments.
The authors are supported by the Yushan Young Scholarship 112V1039 from the Ministry of Education (MOE) of Taiwan, and also by the National Science and Technology Council (NSTC) grant 114L7329. JRN is also supported by NSTC grant 114-2923-M-002-011-MY5.
CHS is also funded by the European Union
under ERC Starting Grant AmpEFT 101165689. 
Views and opinions expressed are however
those of the author(s) only and do not necessarily reflect those of the European Union or the European Research Council. Neither the European Union nor the granting authority can be
held responsible for them.

\vskip .3 cm 
\bibliography{positivity}

@article{Murphy:2020rsh,
    author = "Murphy, Christopher W.",
    title = "{Dimension-8 operators in the Standard Model effective field theory}",
    eprint = "2005.00059",
    archivePrefix = "arXiv",
    primaryClass = "hep-ph",
    doi = "10.1007/JHEP10(2020)174",
    journal = "JHEP",
    volume = "10",
    pages = "174",
    year = "2020"
}

@article{Alonso:2016oah,
    author = "Alonso, Rodrigo and Jenkins, Elizabeth E. and Manohar, Aneesh V.",
    title = "{Geometry of the scalar sector}",
    eprint = "1605.03602",
    archivePrefix = "arXiv",
    primaryClass = "hep-ph",
    reportNumber = "CERN-TH-2016-116",
    doi = "10.1007/JHEP08(2016)101",
    journal = "JHEP",
    volume = "08",
    pages = "101",
    year = "2016"
}

@article{Chala:2023jyx,
    author = "Chala, Mikael",
    title = "{Constraints on anomalous dimensions from the positivity of the S matrix}",
    eprint = "2301.09995",
    archivePrefix = "arXiv",
    primaryClass = "hep-ph",
    doi = "10.1103/PhysRevD.108.015031",
    journal = "Phys. Rev. D",
    volume = "108",
    number = "1",
    pages = "015031",
    year = "2023"
}

@article{Chala:2023xjy,
    author = "Chala, Mikael and Li, Xu",
    title = "{Positivity restrictions on the mixing of dimension-eight SMEFT operators}",
    eprint = "2309.16611",
    archivePrefix = "arXiv",
    primaryClass = "hep-ph",
    doi = "10.1103/PhysRevD.109.065015",
    journal = "Phys. Rev. D",
    volume = "109",
    number = "6",
    pages = "065015",
    year = "2024"
}

@article{Adams:2006sv,
    author = "Adams, Allan and Arkani-Hamed, Nima and Dubovsky, Sergei and Nicolis, Alberto and Rattazzi, Riccardo",
    title = "{Causality, analyticity and an IR obstruction to UV completion}",
    eprint = "hep-th/0602178",
    archivePrefix = "arXiv",
    reportNumber = "CERN-PH-TH-2006-033, HUTP-06-A0005",
    doi = "10.1088/1126-6708/2006/10/014",
    journal = "JHEP",
    volume = "10",
    pages = "014",
    year = "2006"
}

@article{Bern:2019wie,
    author = "Bern, Zvi and Parra-Martinez, Julio and Sawyer, Eric",
    title = "{Nonrenormalization and operator mixing via on-shell methods}",
    eprint = "1910.05831",
    archivePrefix = "arXiv",
    primaryClass = "hep-ph",
    reportNumber = "UCLA/TEP/2019/105, CERN-TH-2019-160",
    doi = "10.1103/PhysRevLett.124.051601",
    journal = "Phys. Rev. Lett.",
    volume = "124",
    number = "5",
    pages = "051601",
    year = "2020"
}

@article{Cao:2023adc,
    author = "Cao, Weiguang and Herzog, Franz and Melia, Tom and Roosmale Nepveu, Jasper",
    title = "{Non-linear non-renormalization theorems}",
    eprint = "2303.07391",
    archivePrefix = "arXiv",
    primaryClass = "hep-ph",
    doi = "10.1007/JHEP08(2023)080",
    journal = "JHEP",
    volume = "08",
    pages = "080",
    year = "2023"
}

@article{Jiang:2020rwz,
    author = "Jiang, Minyuan and Shu, Jing and Xiao, Ming-Lei and Zheng, Yu-Hui",
    title = "{Partial wave amplitude basis and selection rules in effective field theories}",
    eprint = "2001.04481",
    archivePrefix = "arXiv",
    primaryClass = "hep-ph",
    doi = "10.1103/PhysRevLett.126.011601",
    journal = "Phys. Rev. Lett.",
    volume = "126",
    number = "1",
    pages = "011601",
    year = "2021"
}

@article{Machado:2022ozb,
    author = "Machado, Camila S. and Renner, Sophie and Sutherland, Dave",
    title = "{Building blocks of the flavourful SMEFT RG}",
    eprint = "2210.09316",
    archivePrefix = "arXiv",
    primaryClass = "hep-ph",
    reportNumber = "DESY-22-161",
    doi = "10.1007/JHEP03(2023)226",
    journal = "JHEP",
    volume = "03",
    pages = "226",
    year = "2023"
}

@article{Li:2022aby,
    author = "Li, Xu",
    title = "{Positivity bounds at one-loop level: the Higgs sector}",
    eprint = "2212.12227",
    archivePrefix = "arXiv",
    primaryClass = "hep-ph",
    doi = "10.1007/JHEP05(2023)230",
    journal = "JHEP",
    volume = "05",
    pages = "230",
    year = "2023"
}

@article{Brivio_2019,
   title={The Standard Model as an effective field theory},
   volume={793},
   ISSN={0370-1573},
   url={http://dx.doi.org/10.1016/j.physrep.2018.11.002},
   DOI={10.1016/j.physrep.2018.11.002},
   journal={Physics Reports},
   publisher={Elsevier BV},
   author={Brivio, Ilaria and Trott, Michael},
   year={2019},
   month=feb, pages={1–98} }

@article{EliasMiro:2020tdv,
    author = "Elias Mir\'o, Joan and Ingoldby, James and Riembau, Marc",
    title = "{EFT anomalous dimensions from the S-matrix}",
    eprint = "2005.06983",
    archivePrefix = "arXiv",
    primaryClass = "hep-ph",
    doi = "10.1007/JHEP09(2020)163",
    journal = "JHEP",
    volume = "09",
    pages = "163",
    year = "2020"
}

@article{Baratella:2020lzz,
    author = "Baratella, Pietro and Fernandez, Clara and Pomarol, Alex",
    title = "{Renormalization of higher-dimensional operators from on-shell amplitudes}",
    eprint = "2005.07129",
    archivePrefix = "arXiv",
    primaryClass = "hep-ph",
    doi = "10.1016/j.nuclphysb.2020.115155",
    journal = "Nucl. Phys. B",
    volume = "959",
    pages = "115155",
    year = "2020"
}

@article{Bresciani:2024shu,
    author = "Bresciani, Luigi C. and Brunello, Giacomo and Levati, Gabriele and Mastrolia, Pierpaolo and Paradisi, Paride",
    title = "{Renormalization of effective field theories via on-shell methods: the case of axion-like particles}",
    eprint = "2412.04160",
    archivePrefix = "arXiv",
    primaryClass = "hep-ph",
    doi = "10.1007/JHEP10(2025)190",
    journal = "JHEP",
    volume = "10",
    pages = "190",
    year = "2025"
}

@article{Bresciani:2023jsu,
    author = "Bresciani, L. C. and Levati, G. and Mastrolia, P. and Paradisi, P.",
    title = "{Anomalous dimensions via on-shell methods: operator mixing and leading mass effects}",
    eprint = "2312.05206",
    archivePrefix = "arXiv",
    primaryClass = "hep-ph",
    doi = "10.1103/PhysRevD.110.056041",
    journal = "Phys. Rev. D",
    volume = "110",
    number = "5",
    pages = "056041",
    year = "2024"
}

@article{Davidson:2018zuo,
    author = "Davidson, Sacha and Gorbahn, Martin and Leak, Matthew",
    title = "{Majorana neutrino masses in the renormalization group equations for lepton flavor violation}",
    eprint = "1807.04283",
    archivePrefix = "arXiv",
    primaryClass = "hep-ph",
    doi = "10.1103/PhysRevD.98.095014",
    journal = "Phys. Rev. D",
    volume = "98",
    number = "9",
    pages = "095014",
    year = "2018"
}

@article{Weinberg:1979sa,
    author = "Weinberg, Steven",
    title = "{Baryon and lepton nonconserving processes}",
    reportNumber = "HUTP-79-A050",
    doi = "10.1103/PhysRevLett.43.1566",
    journal = "Phys. Rev. Lett.",
    volume = "43",
    pages = "1566--1570",
    year = "1979"
}

@article{Broncano:2004tz,
    author = "Broncano, A. and Gavela, M. B. and Jenkins, Elizabeth Ellen",
    title = "{Renormalization of lepton mixing for Majorana neutrinos}",
    eprint = "hep-ph/0406019",
    archivePrefix = "arXiv",
    reportNumber = "FTUAM-04-11, IFT-UAM-CSIC-04-27, UCSD-PTH-04-07",
    doi = "10.1016/j.nuclphysb.2004.11.001",
    journal = "Nucl. Phys. B",
    volume = "705",
    pages = "269--295",
    year = "2005"
}

@article{Grzadkowski:2010es,
    author = "Grzadkowski, B. and Iskrzynski, M. and Misiak, M. and Rosiek, J.",
    title = "{Dimension-Six Terms in the Standard Model Lagrangian}",
    eprint = "1008.4884",
    archivePrefix = "arXiv",
    primaryClass = "hep-ph",
    reportNumber = "IFT-9-2010, TTP10-35",
    doi = "10.1007/JHEP10(2010)085",
    journal = "JHEP",
    volume = "10",
    pages = "085",
    year = "2010"
}

@article{AccettulliHuber:2021uoa,
    author = "Accettulli Huber, Manuel and De Angelis, Stefano",
    title = "{Standard Model EFTs via on-shell methods}",
    eprint = "2108.03669",
    archivePrefix = "arXiv",
    primaryClass = "hep-th",
    reportNumber = "QMUL-PH-21-32, SAGEX-21-17-E",
    doi = "10.1007/JHEP11(2021)221",
    journal = "JHEP",
    volume = "11",
    pages = "221",
    year = "2021"
}

@article{PhysRevLett.30.1343,
  title = {Ultraviolet Behavior of Non-Abelian Gauge Theories},
  author = {Gross, David J. and Wilczek, Frank},
  journal = {Phys. Rev. Lett.},
  volume = {30},
  issue = {26},
  pages = {1343--1346},
  numpages = {0},
  year = {1973},
  month = {Jun},
  publisher = {American Physical Society},
  doi = {10.1103/PhysRevLett.30.1343},
  url = {https://link.aps.org/doi/10.1103/PhysRevLett.30.1343}
}

@article{PhysRevLett.30.1346,
  title = {Reliable Perturbative Results for Strong Interactions?},
  author = {Politzer, H. David},
  journal = {Phys. Rev. Lett.},
  volume = {30},
  issue = {26},
  pages = {1346--1349},
  numpages = {0},
  year = {1973},
  month = {Jun},
  publisher = {American Physical Society},
  doi = {10.1103/PhysRevLett.30.1346},
  url = {https://link.aps.org/doi/10.1103/PhysRevLett.30.1346}
}

@article{Jenkins:2023bls,
    author = "Jenkins, Elizabeth E. and Manohar, Aneesh V. and Naterop, Luca and Pag\`es, Julie",
    title = "{Two loop renormalization of scalar theories using a geometric approach}",
    eprint = "2310.19883",
    archivePrefix = "arXiv",
    primaryClass = "hep-ph",
    reportNumber = "ZU-TH 69/23, PSI-PR-23-39",
    doi = "10.1007/JHEP02(2024)131",
    journal = "JHEP",
    volume = "02",
    pages = "131",
    year = "2024"
}

@article{Zamolodchikov:1986gt,
    author = "Zamolodchikov, A. B.",
    title = "{Irreversibility of the Flux of the Renormalization Group in a 2D Field Theory}",
    journal = "JETP Lett.",
    volume = "43",
    pages = "730--732",
    year = "1986"
}

@article{Cardy:1988cwa,
    author = "Cardy, John L.",
    title = "{Is There a c Theorem in Four-Dimensions?}",
    doi = "10.1016/0370-2693(88)90054-8",
    journal = "Phys. Lett. B",
    volume = "215",
    pages = "749--752",
    year = "1988"
}

@article{Komargodski:2011vj,
    author = "Komargodski, Zohar and Schwimmer, Adam",
    title = "{On Renormalization Group Flows in Four Dimensions}",
    eprint = "1107.3987",
    archivePrefix = "arXiv",
    primaryClass = "hep-th",
    doi = "10.1007/JHEP12(2011)099",
    journal = "JHEP",
    volume = "12",
    pages = "099",
    year = "2011"
}

@article{Komargodski:2011xv,
    author = "Komargodski, Zohar",
    title = "{The Constraints of Conformal Symmetry on RG Flows}",
    eprint = "1112.4538",
    archivePrefix = "arXiv",
    primaryClass = "hep-th",
    reportNumber = "WIS-12-11-DEC-DPPA",
    doi = "10.1007/JHEP07(2012)069",
    journal = "JHEP",
    volume = "07",
    pages = "069",
    year = "2012"
}

@article{Luty:2012ww,
    author = "Luty, Markus A. and Polchinski, Joseph and Rattazzi, Riccardo",
    title = "{The $a$-theorem and the Asymptotics of 4D Quantum Field Theory}",
    eprint = "1204.5221",
    archivePrefix = "arXiv",
    primaryClass = "hep-th",
    doi = "10.1007/JHEP01(2013)152",
    journal = "JHEP",
    volume = "01",
    pages = "152",
    year = "2013"
}

@article{Henning:2015alf,
    author = "Henning, Brian and Lu, Xiaochuan and Melia, Tom and Murayama, Hitoshi",
    title = "{2, 84, 30, 993, 560, 15456, 11962, 261485, ...: higher dimension operators in the SM EFT}",
    eprint = "1512.03433",
    archivePrefix = "arXiv",
    primaryClass = "hep-ph",
    reportNumber = "UCB-PTH-15-14, IPMU15-0207",
    doi = "10.1007/JHEP08(2017)016",
    journal = "JHEP",
    volume = "08",
    pages = "016",
    year = "2017",
    note = "[Erratum: JHEP 09, 019 (2019)]"
}

@article{Cronin:1967jq,
    author = "Cronin, Jeremiah A.",
    title = "{Phenomenological model of strong and weak interactions in chiral U(3) x U(3)}",
    doi = "10.1103/PhysRev.161.1483",
    journal = "Phys. Rev.",
    volume = "161",
    pages = "1483--1494",
    year = "1967"
}

@article{Weinberg:1966fm,
    author = "Weinberg, Steven",
    title = "{Dynamical approach to current algebra}",
    doi = "10.1103/PhysRevLett.18.188",
    journal = "Phys. Rev. Lett.",
    volume = "18",
    pages = "188--191",
    year = "1967"
}

@article{Weinberg:1968de,
    author = "Weinberg, Steven",
    title = "{Nonlinear realizations of chiral symmetry}",
    doi = "10.1103/PhysRev.166.1568",
    journal = "Phys. Rev.",
    volume = "166",
    pages = "1568--1577",
    year = "1968"
}

@article{Alonso:2014rga,
    author = "Alonso, Rodrigo and Jenkins, Elizabeth E. and Manohar, Aneesh V.",
    title = "{Holomorphy without supersymmetry in the Standard Model effective field theory}",
    eprint = "1409.0868",
    archivePrefix = "arXiv",
    primaryClass = "hep-ph",
    doi = "10.1016/j.physletb.2014.10.045",
    journal = "Phys. Lett. B",
    volume = "739",
    pages = "95--98",
    year = "2014"
}

@article{Elias-Miro:2014eia,
    author = "Elias-Miro, J. and Espinosa, J. R. and Pomarol, A.",
    title = "{One-loop non-renormalization results in EFTs}",
    eprint = "1412.7151",
    archivePrefix = "arXiv",
    primaryClass = "hep-ph",
    doi = "10.1016/j.physletb.2015.05.056",
    journal = "Phys. Lett. B",
    volume = "747",
    pages = "272--280",
    year = "2015"
}

@article{Cheung:2015aba,
    author = "Cheung, Clifford and Shen, Chia-Hsien",
    title = "{Nonrenormalization theorems without supersymmetry}",
    eprint = "1505.01844",
    archivePrefix = "arXiv",
    primaryClass = "hep-ph",
    reportNumber = "CALT-2015-024",
    doi = "10.1103/PhysRevLett.115.071601",
    journal = "Phys. Rev. Lett.",
    volume = "115",
    number = "7",
    pages = "071601",
    year = "2015"
}

@article{EliasMiro:2021jgu,
    author = "Elias Miro, Joan and Fernandez, Clara and Gumus, Mehmet Asim and Pomarol, Alex",
    title = "{Gearing up for the next generation of LFV experiments, via on-shell methods}",
    eprint = "2112.12131",
    archivePrefix = "arXiv",
    primaryClass = "hep-ph",
    doi = "10.1007/JHEP06(2022)126",
    journal = "JHEP",
    volume = "06",
    pages = "126",
    year = "2022"
}

@article{Caron-Huot:2016cwu,
    author = "Caron-Huot, Simon and Wilhelm, Matthias",
    title = "{Renormalization group coefficients and the S-matrix}",
    eprint = "1607.06448",
    archivePrefix = "arXiv",
    primaryClass = "hep-th",
    doi = "10.1007/JHEP12(2016)010",
    journal = "JHEP",
    volume = "12",
    pages = "010",
    year = "2016"
}

@article{Bern:2020ikv,
    author = "Bern, Zvi and Parra-Martinez, Julio and Sawyer, Eric",
    title = "{Structure of two-loop SMEFT anomalous dimensions via on-shell methods}",
    eprint = "2005.12917",
    archivePrefix = "arXiv",
    primaryClass = "hep-ph",
    doi = "10.1007/JHEP10(2020)211",
    journal = "JHEP",
    volume = "10",
    pages = "211",
    year = "2020"
}

@article{Jiang:2020mhe,
    author = "Jiang, Minyuan and Ma, Teng and Shu, Jing",
    title = "{Renormalization group evolution from on-shell SMEFT}",
    eprint = "2005.10261",
    archivePrefix = "arXiv",
    primaryClass = "hep-ph",
    doi = "10.1007/JHEP01(2021)101",
    journal = "JHEP",
    volume = "01",
    pages = "101",
    year = "2021"
}

@article{Baratella:2020dvw,
    author = "Baratella, Pietro and Fernandez, Clara and von Harling, Benedict and Pomarol, Alex",
    title = "{Anomalous dimensions of dffective theories from partial waves}",
    eprint = "2010.13809",
    archivePrefix = "arXiv",
    primaryClass = "hep-ph",
    reportNumber = "TUM-HEP-1291/20",
    doi = "10.1007/JHEP03(2021)287",
    journal = "JHEP",
    volume = "03",
    pages = "287",
    year = "2021"
}

@article{Baratella:2021guc,
    author = "Baratella, Pietro and Haslehner, Dominik and Ruhdorfer, Maximilian and Serra, Javi and Weiler, Andreas",
    title = "{RG of GR from on-shell amplitudes}",
    eprint = "2109.06191",
    archivePrefix = "arXiv",
    primaryClass = "hep-th",
    reportNumber = "TUM-HEP-1363/21",
    doi = "10.1007/JHEP03(2022)156",
    journal = "JHEP",
    volume = "03",
    pages = "156",
    year = "2022"
}

@article{Baratella:2022nog,
    author = "Baratella, Pietro and Maggio, Sara and Stadlbauer, Michael and Theil, Tobias",
    title = "{Two-loop infrared renormalization with on-shell methods}",
    eprint = "2207.08831",
    archivePrefix = "arXiv",
    primaryClass = "hep-th",
    reportNumber = "TUM-HEP-1410/22, BONN-TH-2023-01",
    doi = "10.1140/epjc/s10052-023-11929-6",
    journal = "Eur. Phys. J. C",
    volume = "83",
    number = "8",
    pages = "751",
    year = "2023"
}

@article{Ye:2024rzr,
    author = "Ye, Yunxiao and He, Bin and Gu, Jiayin",
    title = "{Positivity bounds in scalar effective field theories at one-loop level}",
    eprint = "2408.10318",
    archivePrefix = "arXiv",
    primaryClass = "hep-ph",
    doi = "10.1007/JHEP12(2024)046",
    journal = "JHEP",
    volume = "12",
    pages = "046",
    year = "2024"
}

@article{Manohar:1983md,
    author = "Manohar, Aneesh and Georgi, Howard",
    title = "{Chiral quarks and the nonrelativistic quark model}",
    reportNumber = "HUTP-83/A042a",
    doi = "10.1016/0550-3213(84)90231-1",
    journal = "Nucl. Phys. B",
    volume = "234",
    pages = "189--212",
    year = "1984"
}

@article{Jenkins:2013zja,
    author = "Jenkins, Elizabeth E. and Manohar, Aneesh V. and Trott, Michael",
    title = "{Renormalization group evolution of the Standard Model dimension six operators I: formalism and lambda dependence}",
    eprint = "1308.2627",
    archivePrefix = "arXiv",
    primaryClass = "hep-ph",
    doi = "10.1007/JHEP10(2013)087",
    journal = "JHEP",
    volume = "10",
    pages = "087",
    year = "2013"
}

@article{Jenkins:2013wua,
    author = "Jenkins, Elizabeth E. and Manohar, Aneesh V. and Trott, Michael",
    title = "{Renormalization group evolution of the Standard Model dimension six operators II: Yukawa dependence}",
    eprint = "1310.4838",
    archivePrefix = "arXiv",
    primaryClass = "hep-ph",
    reportNumber = "CERN-PH-TH/2015-247",
    doi = "10.1007/JHEP01(2014)035",
    journal = "JHEP",
    volume = "01",
    pages = "035",
    year = "2014"
}

@article{Alonso:2013hga,
    author = "Alonso, Rodrigo and Jenkins, Elizabeth E. and Manohar, Aneesh V. and Trott, Michael",
    title = "{Renormalization group evolution of the Standard Model dimension six operators III: gauge coupling dependence and phenomenology}",
    eprint = "1312.2014",
    archivePrefix = "arXiv",
    primaryClass = "hep-ph",
    reportNumber = "CERN-PH-TH-2013-305, CERN-PH-TH/2013-305",
    doi = "10.1007/JHEP04(2014)159",
    journal = "JHEP",
    volume = "04",
    pages = "159",
    year = "2014"
}

@article{Alonso:2014zka,
    author = "Alonso, Rodrigo and Chang, Hsi-Ming and Jenkins, Elizabeth E. and Manohar, Aneesh V. and Shotwell, Brian",
    title = "{Renormalization group evolution of dimension-six baryon number violating operators}",
    eprint = "1405.0486",
    archivePrefix = "arXiv",
    primaryClass = "hep-ph",
    doi = "10.1016/j.physletb.2014.05.065",
    journal = "Phys. Lett. B",
    volume = "734",
    pages = "302--307",
    year = "2014"
}

@article{Chala:2021pll,
    author = "Chala, Mikael and Guedes, Guilherme and Ramos, Maria and Santiago, Jose",
    title = "{Towards the renormalisation of the Standard Model effective field theory to dimension eight: bosonic interactions I}",
    eprint = "2106.05291",
    archivePrefix = "arXiv",
    primaryClass = "hep-ph",
    doi = "10.21468/SciPostPhys.11.3.065",
    journal = "SciPost Phys.",
    volume = "11",
    pages = "065",
    year = "2021"
}

@article{DasBakshi:2022mwk,
    author = "Das Bakshi, Supratim and Chala, Mikael and D\'\i{}az-Carmona, \'Alvaro and Guedes, Guilherme",
    title = "{Towards the renormalisation of the Standard Model effective field theory to dimension eight: bosonic interactions II}",
    eprint = "2205.03301",
    archivePrefix = "arXiv",
    primaryClass = "hep-ph",
    doi = "10.1140/epjp/s13360-022-03194-5",
    journal = "Eur. Phys. J. Plus",
    volume = "137",
    number = "8",
    pages = "973",
    year = "2022"
}

@article{Bakshi:2024wzz,
    author = "Bakshi, S. D. and Chala, M. and D\'\i{}az-Carmona, \'A. and Ren, Z. and Vilches, F.",
    title = "{Renormalization of the SMEFT to dimension eight: fermionic interactions I}",
    eprint = "2409.15408",
    archivePrefix = "arXiv",
    primaryClass = "hep-ph",
    doi = "10.1007/JHEP12(2024)214",
    journal = "JHEP",
    volume = "12",
    pages = "214",
    year = "2025"
}

@article{Boughezal:2024zqa,
    author = "Boughezal, Radja and Huang, Yingsheng and Petriello, Frank",
    title = "{Renormalization-group running of dimension-8 four-fermion operators in the SMEFT}",
    eprint = "2408.15378",
    archivePrefix = "arXiv",
    primaryClass = "hep-ph",
    doi = "10.1103/PhysRevD.110.116015",
    journal = "Phys. Rev. D",
    volume = "110",
    number = "11",
    pages = "116015",
    year = "2024"
}

@article{Helset:2022tlf,
    author = "Helset, Andreas and Jenkins, Elizabeth E. and Manohar, Aneesh V.",
    title = "{Geometry in scattering amplitudes}",
    eprint = "2210.08000",
    archivePrefix = "arXiv",
    primaryClass = "hep-ph",
    reportNumber = "CALT-TH-2022-036",
    doi = "10.1103/PhysRevD.106.116018",
    journal = "Phys. Rev. D",
    volume = "106",
    number = "11",
    pages = "116018",
    year = "2022"
}

@article{Helset:2022pde,
    author = "Helset, Andreas and Jenkins, Elizabeth E. and Manohar, Aneesh V.",
    title = "{Renormalization of the Standard Model effective field theory from geometry}",
    eprint = "2212.03253",
    archivePrefix = "arXiv",
    primaryClass = "hep-ph",
    reportNumber = "CALT-TH-2022-041",
    doi = "10.1007/JHEP02(2023)063",
    journal = "JHEP",
    volume = "02",
    pages = "063",
    year = "2023"
}

@article{Assi:2023zid,
    author = "Assi, Beno\^\i{}t and Helset, Andreas and Manohar, Aneesh V. and Pag\`es, Julie and Shen, Chia-Hsien",
    title = "{Fermion geometry and the renormalization of the Standard Model Effective Field Theory}",
    eprint = "2307.03187",
    archivePrefix = "arXiv",
    primaryClass = "hep-ph",
    reportNumber = "CALT-TH-2023-024, FERMILAB-PUB-23-362-T",
    doi = "10.1007/JHEP11(2023)201",
    journal = "JHEP",
    volume = "11",
    pages = "201",
    year = "2023"
}

@article{Pham:1985cr,
    author = "Pham, T. N. and Truong, Tran N.",
    title = "{Evaluation of the derivative quartic terms of the meson chiral Lagrangian from forward dispersion relation}",
    reportNumber = "Print-85-0588 (ECOLE POLY)",
    doi = "10.1103/PhysRevD.31.3027",
    journal = "Phys. Rev. D",
    volume = "31",
    pages = "3027",
    year = "1985"
}

@article{Remmen:2019cyz,
    author = "Remmen, Grant N. and Rodd, Nicholas L.",
    title = "{Consistency of the Standard Model effective field theory}",
    eprint = "1908.09845",
    archivePrefix = "arXiv",
    primaryClass = "hep-ph",
    doi = "10.1007/JHEP12(2019)032",
    journal = "JHEP",
    volume = "12",
    pages = "032",
    year = "2019"
}

@article{Remmen:2020vts,
    author = "Remmen, Grant N. and Rodd, Nicholas L.",
    title = "{Flavor constraints from unitarity and analyticity}",
    eprint = "2004.02885",
    archivePrefix = "arXiv",
    primaryClass = "hep-ph",
    doi = "10.1103/PhysRevLett.127.149901",
    journal = "Phys. Rev. Lett.",
    volume = "125",
    number = "8",
    pages = "081601",
    year = "2020",
    note = "[Erratum: Phys.Rev.Lett. 127, 149901 (2021)]"
}

@article{Remmen:2022orj,
    author = "Remmen, Grant N. and Rodd, Nicholas L.",
    title = "{Spinning sum rules for the dimension-six SMEFT}",
    eprint = "2206.13524",
    archivePrefix = "arXiv",
    primaryClass = "hep-ph",
    reportNumber = "CERN-TH-2022-105",
    doi = "10.1007/JHEP09(2022)030",
    journal = "JHEP",
    volume = "09",
    pages = "030",
    year = "2022"
}

@article{Zhang:2018shp,
    author = "Zhang, Cen and Zhou, Shuang-Yong",
    title = "{Positivity bounds on vector boson scattering at the LHC}",
    eprint = "1808.00010",
    archivePrefix = "arXiv",
    primaryClass = "hep-ph",
    reportNumber = "USTC-ICTS-18-13",
    doi = "10.1103/PhysRevD.100.095003",
    journal = "Phys. Rev. D",
    volume = "100",
    number = "9",
    pages = "095003",
    year = "2019"
}

@article{Remmen:2020uze,
    author = "Remmen, Grant N. and Rodd, Nicholas L.",
    title = "{Signs, spin, SMEFT: sum rules at dimension six}",
    eprint = "2010.04723",
    archivePrefix = "arXiv",
    primaryClass = "hep-ph",
    doi = "10.1103/PhysRevD.105.036006",
    journal = "Phys. Rev. D",
    volume = "105",
    number = "3",
    pages = "036006",
    year = "2022"
}

@article{Low:2009di,
    author = "Low, Ian and Rattazzi, Riccardo and Vichi, Alessandro",
    title = "{Theoretical constraints on the Higgs effective couplings}",
    eprint = "0907.5413",
    archivePrefix = "arXiv",
    primaryClass = "hep-ph",
    doi = "10.1007/JHEP04(2010)126",
    journal = "JHEP",
    volume = "04",
    pages = "126",
    year = "2010"
}

@article{Zhang:2020jyn,
    author = "Zhang, Cen and Zhou, Shuang-Yong",
    title = "{Convex geometry perspective on the (Standard Model) effective field theory space}",
    eprint = "2005.03047",
    archivePrefix = "arXiv",
    primaryClass = "hep-ph",
    reportNumber = "USTC-ICTS/PCFT-20-14",
    doi = "10.1103/PhysRevLett.125.201601",
    journal = "Phys. Rev. Lett.",
    volume = "125",
    number = "20",
    pages = "201601",
    year = "2020"
}

@article{Yamashita:2020gtt,
    author = "Yamashita, Kimiko and Zhang, Cen and Zhou, Shuang-Yong",
    title = "{Elastic positivity vs extremal positivity bounds in SMEFT: a case study in transversal electroweak gauge-boson scatterings}",
    eprint = "2009.04490",
    archivePrefix = "arXiv",
    primaryClass = "hep-ph",
    reportNumber = "USTC-ICTS/PCFT-20-29",
    doi = "10.1007/JHEP01(2021)095",
    journal = "JHEP",
    volume = "01",
    pages = "095",
    year = "2021"
}

@inproceedings{Hong:2024fbl,
    author = "Hong, Dong-Yu and Wang, Zhuo-Hui and Zhou, Shuang-Yong",
    title = "{On Capped Higgs positivity cone}",
    eprint = "2404.04479",
    archivePrefix = "arXiv",
    primaryClass = "hep-ph",
    month = "4",
    year = "2024"
}

@article{Chen:2023bhu,
    author = "Chen, Qing and Mimasu, Ken and Wu, Tong Arthur and Zhang, Guo-Dong and Zhou, Shuang-Yong",
    title = "{Capping the positivity cone: dimension-8 Higgs operators in the SMEFT}",
    eprint = "2309.15922",
    archivePrefix = "arXiv",
    primaryClass = "hep-ph",
    doi = "10.1007/JHEP03(2024)180",
    journal = "JHEP",
    volume = "03",
    pages = "180",
    year = "2024"
}

@article{Remmen:2024hry,
    author = "Remmen, Grant N. and Rodd, Nicholas L.",
    title = "{Positively identifying Higgs effective field theory or standard model effective field theory}",
    eprint = "2412.07827",
    archivePrefix = "arXiv",
    primaryClass = "hep-ph",
    doi = "10.1103/vj7j-zj11",
    journal = "Phys. Rev. D",
    volume = "113",
    number = "3",
    pages = "036027",
    year = "2026"
}

@article{Arkani-Hamed:2020blm,
    author = "Arkani-Hamed, Nima and Huang, Tzu-Chen and Huang, Yu-tin",
    title = "{The EFT-Hedron}",
    eprint = "2012.15849",
    archivePrefix = "arXiv",
    primaryClass = "hep-th",
    reportNumber = "NCTS-TH/2014, CALT-TH 2020-061",
    doi = "10.1007/JHEP05(2021)259",
    journal = "JHEP",
    volume = "05",
    pages = "259",
    year = "2021"
}

@article{Bellazzini:2020cot,
    author = "Bellazzini, Brando and Elias Mir\'o, Joan and Rattazzi, Riccardo and Riembau, Marc and Riva, Francesco",
    title = "{Positive moments for scattering amplitudes}",
    eprint = "2011.00037",
    archivePrefix = "arXiv",
    primaryClass = "hep-th",
    doi = "10.1103/PhysRevD.104.036006",
    journal = "Phys. Rev. D",
    volume = "104",
    number = "3",
    pages = "036006",
    year = "2021"
}

@article{Bellazzini:2021oaj,
    author = "Bellazzini, Brando and Riembau, Marc and Riva, Francesco",
    title = "{IR side of positivity bounds}",
    eprint = "2112.12561",
    archivePrefix = "arXiv",
    primaryClass = "hep-th",
    doi = "10.1103/PhysRevD.106.105008",
    journal = "Phys. Rev. D",
    volume = "106",
    number = "10",
    pages = "105008",
    year = "2022"
}

@article{Beadle:2024hqg,
    author = "Beadle, Carl and Isabella, Giulia and Perrone, Davide and Ricossa, Sara and Riva, Francesco and Serra, Francesco",
    title = "{Non-forward UV/IR relations}",
    eprint = "2407.02346",
    archivePrefix = "arXiv",
    primaryClass = "hep-th",
    doi = "10.1007/JHEP08(2025)188",
    journal = "JHEP",
    volume = "08",
    pages = "188",
    year = "2025"
}

@article{Beadle:2025cdx,
    author = "Beadle, Carl and Isabella, Giulia and Perrone, Davide and Ricossa, Sara and Riva, Francesco and Serra, Francesco",
    title = "{The EFT bootstrap at finite M$_{PL}$}",
    eprint = "2501.18465",
    archivePrefix = "arXiv",
    primaryClass = "hep-th",
    doi = "10.1007/JHEP06(2025)209",
    journal = "JHEP",
    volume = "06",
    pages = "209",
    year = "2025"
}

@article{Desai:2025alt,
    author = "Desai, Jay and Ghosh, Diptimoy",
    title = "{Positivity at 1-loop: bounds on photon and gluon EFTs}",
    eprint = "2503.21864",
    archivePrefix = "arXiv",
    primaryClass = "hep-ph",
    doi = "10.1007/JHEP09(2025)152",
    journal = "JHEP",
    volume = "09",
    pages = "152",
    year = "2025"
}

@article{Chang:2025cxc,
    author = "Chang, Cyuan-Han and Parra-Martinez, Julio",
    title = "{Graviton loops and negativity}",
    eprint = "2501.17949",
    archivePrefix = "arXiv",
    primaryClass = "hep-th",
    doi = "10.1007/JHEP08(2025)175",
    journal = "JHEP",
    volume = "08",
    pages = "175",
    year = "2025"
}

@article{Chala:2021wpj,
    author = "Chala, Mikael and Santiago, Jose",
    title = "{Positivity bounds in the Standard Model effective field theory beyond tree level}",
    eprint = "2110.01624",
    archivePrefix = "arXiv",
    primaryClass = "hep-ph",
    doi = "10.1103/PhysRevD.105.L111901",
    journal = "Phys. Rev. D",
    volume = "105",
    number = "11",
    pages = "L111901",
    year = "2022"
}

@article{Bi:2019phv,
    author = "Bi, Qi and Zhang, Cen and Zhou, Shuang-Yong",
    title = "{Positivity constraints on aQGC: carving out the physical parameter space}",
    eprint = "1902.08977",
    archivePrefix = "arXiv",
    primaryClass = "hep-ph",
    reportNumber = "USTC-ICTS-19-01",
    doi = "10.1007/JHEP06(2019)137",
    journal = "JHEP",
    volume = "06",
    pages = "137",
    year = "2019"
}

@article{Assi:2024zap,
    author = "Assi, Beno\^\i{}t and Martin, Adam",
    title = "{Energy-enhanced dimension eight SMEFT effects in VBF Higgs production}",
    eprint = "2410.21563",
    archivePrefix = "arXiv",
    primaryClass = "hep-ph",
    reportNumber = "FERMILAB-PUB-24-0786-V",
    doi = "10.1007/JHEP02(2025)029",
    journal = "JHEP",
    volume = "02",
    pages = "029",
    year = "2025"
}

@article{Grojean:2024tcw,
    author = "Grojean, Christophe and Guedes, Guilherme and Roosmale Nepveu, Jasper and Salla, Gabriel M.",
    title = "{A log story short: running contributions to radiative Higgs decays in the SMEFT}",
    eprint = "2405.20371",
    archivePrefix = "arXiv",
    primaryClass = "hep-ph",
    reportNumber = "CERN-TH-2024-075, DESY-24-077, HU-EP-24/15-RTG",
    doi = "10.1007/JHEP12(2024)065",
    journal = "JHEP",
    volume = "12",
    pages = "065",
    year = "2024"
}

@article{Boughezal:2022nof,
    author = "Boughezal, Radja and Huang, Yingsheng and Petriello, Frank",
    title = "{Exploring the SMEFT at dimension eight with Drell-Yan transverse momentum measurements}",
    eprint = "2207.01703",
    archivePrefix = "arXiv",
    primaryClass = "hep-ph",
    doi = "10.1103/PhysRevD.106.036020",
    journal = "Phys. Rev. D",
    volume = "106",
    number = "3",
    pages = "036020",
    year = "2022"
}

@article{DasBakshi:2024krs,
    author = "Das Bakshi, Supratim and Dawson, Sally and Fontes, Duarte and Homiller, Samuel",
    title = "{Relevance of one-loop SMEFT matching in the 2HDM}",
    eprint = "2401.12279",
    archivePrefix = "arXiv",
    primaryClass = "hep-ph",
    doi = "10.1103/PhysRevD.109.075022",
    journal = "Phys. Rev. D",
    volume = "109",
    number = "7",
    pages = "075022",
    year = "2024"
}

@article{Contino:2016jqw,
    author = "Contino, Roberto and Falkowski, Adam and Goertz, Florian and Grojean, Christophe and Riva, Francesco",
    title = "{On the validity of the effective field theory approach to SM precision tests}",
    eprint = "1604.06444",
    archivePrefix = "arXiv",
    primaryClass = "hep-ph",
    reportNumber = "DESY-16-067, CERN-TH-2016-082, LPT-Orsay-16-32",
    doi = "10.1007/JHEP07(2016)144",
    journal = "JHEP",
    volume = "07",
    pages = "144",
    year = "2016"
}

@article{Biekotter:2025nln,
    author = {Biek{\"o}tter, Anke and Pecjak, Benjamin D.},
    title = "{Analytic results for electroweak precision observables at NLO in SMEFT}",
    eprint = "2503.07724",
    archivePrefix = "arXiv",
    primaryClass = "hep-ph",
    reportNumber = "IPPP/25/13, KIT-TP-04-2025, P3H-25-015",
    doi = "10.1007/JHEP07(2025)134",
    journal = "JHEP",
    volume = "07",
    pages = "134",
    year = "2025"
}

@article{Stefanek:2024kds,
    author = "Stefanek, Ben A.",
    title = "{Non-universal probes of composite Higgs models: new bounds and prospects for FCC-ee}",
    eprint = "2407.09593",
    archivePrefix = "arXiv",
    primaryClass = "hep-ph",
    reportNumber = "KCL-PH-TH/2024-43",
    doi = "10.1007/JHEP09(2024)103",
    journal = "JHEP",
    volume = "09",
    pages = "103",
    year = "2024"
}

@article{Giudice:2007fh,
    author = "Giudice, G. F. and Grojean, C. and Pomarol, A. and Rattazzi, R.",
    title = "{The Strongly-Interacting Light Higgs}",
    eprint = "hep-ph/0703164",
    archivePrefix = "arXiv",
    reportNumber = "CERN-PH-TH-2007-47",
    doi = "10.1088/1126-6708/2007/06/045",
    journal = "JHEP",
    volume = "06",
    pages = "045",
    year = "2007"
}

@article{Manohar:2008tc,
    author = "Manohar, Aneesh V. and Mateu, Vicent",
    title = "{Dispersion Relation Bounds for pi pi Scattering}",
    eprint = "0801.3222",
    archivePrefix = "arXiv",
    primaryClass = "hep-ph",
    reportNumber = "IFIC-08-01, FTUV-07-0121",
    doi = "10.1103/PhysRevD.77.094019",
    journal = "Phys. Rev. D",
    volume = "77",
    pages = "094019",
    year = "2008"
}

@article{DeAngelis:2023bmd,
    author = "De Angelis, Stefano and Durieux, Gauthier",
    title = "{EFT matching from analyticity and unitarity}",
    eprint = "2308.00035",
    archivePrefix = "arXiv",
    primaryClass = "hep-ph",
    reportNumber = "CERN-TH-2023-150",
    doi = "10.21468/SciPostPhys.16.3.071",
    journal = "SciPost Phys.",
    volume = "16",
    pages = "071",
    year = "2024"
}

@article{Alioli:2020kez,
    author = "Alioli, Simone and Boughezal, Radja and Mereghetti, Emanuele and Petriello, Frank",
    title = "{Novel angular dependence in Drell-Yan lepton production via dimension-8 operators}",
    eprint = "2003.11615",
    archivePrefix = "arXiv",
    primaryClass = "hep-ph",
    reportNumber = "LA-UR-20-22498",
    doi = "10.1016/j.physletb.2020.135703",
    journal = "Phys. Lett. B",
    volume = "809",
    pages = "135703",
    year = "2020"
}

@article{Liao:2024xel,
    author = "Liao, Yi and Ma, Xiao-Dong and Wang, Hao-Lin",
    title = "{Probing dimension-8 SMEFT operators through neutral meson mixing}",
    eprint = "2409.10305",
    archivePrefix = "arXiv",
    primaryClass = "hep-ph",
    doi = "10.1007/JHEP03(2025)133",
    journal = "JHEP",
    volume = "03",
    pages = "133",
    year = "2025"
}

@article{Azatov:2016sqh,
    author = "Azatov, Aleksandr and Contino, Roberto and Machado, Camila S. and Riva, Francesco",
    title = "{Positivity constraints on aQGC for BSM amplitudes}",
    eprint = "1607.05236",
    archivePrefix = "arXiv",
    primaryClass = "hep-ph",
    reportNumber = "CERN-TH-2016-165",
    doi = "10.1103/PhysRevD.95.065014",
    journal = "Phys. Rev. D",
    volume = "95",
    number = "6",
    pages = "065014",
    year = "2017"
}

@article{Panico:2018hal,
    author = "Panico, Giuliano and Pomarol, Alex and Riembau, Marc",
    title = "{EFT approach to the electron Electric Dipole Moment at the two-loop level}",
    eprint = "1810.09413",
    archivePrefix = "arXiv",
    primaryClass = "hep-ph",
    reportNumber = "DESY-18-185",
    doi = "10.1007/JHEP04(2019)090",
    journal = "JHEP",
    volume = "04",
    pages = "090",
    year = "2019"
}

@article{Heinrich:2022gzl,
    author = "Heinrich, Gudrun and Lang, Jannis and Scyboz, Ludovic",
    title = "{Beyond dimension six in SM Effective Field Theory: a case study in Higgs pair production at NLO QCD}",
    eprint = "2207.08790",
    archivePrefix = "arXiv",
    primaryClass = "hep-ph",
    reportNumber = "KA-TP-21-2022, P3H-22-079",
    doi = "10.22323/1.416.0009",
    journal = "PoS",
    volume = "LL2022",
    pages = "009",
    year = "2022"
}

@article{Biekotter:2014gup,
    author = {Biek\"otter, Anke and Knochel, Alexander and Kr\"amer, Michael and Liu, Da and Riva, Francesco},
    title = "{Vices and virtues of Higgs effective field theories at large energy}",
    eprint = "1406.7320",
    archivePrefix = "arXiv",
    primaryClass = "hep-ph",
    doi = "10.1103/PhysRevD.91.055029",
    journal = "Phys. Rev. D",
    volume = "91",
    pages = "055029",
    year = "2015"
}

@article{Degrande:2020evl,
    author = "Degrande, C\'eline and Durieux, Gauthier and Maltoni, Fabio and Mimasu, Ken and Vryonidou, Eleni and Zhang, Cen",
    title = "{Automated one-loop computations in the Standard Model effective field theory}",
    eprint = "2008.11743",
    archivePrefix = "arXiv",
    primaryClass = "hep-ph",
    reportNumber = "CERN-TH-2020-140, CP3-20-42",
    doi = "10.1103/PhysRevD.103.096024",
    journal = "Phys. Rev. D",
    volume = "103",
    number = "9",
    pages = "096024",
    year = "2021"
}

@article{Gu:2020ldn,
    author = "Gu, Jiayin and Wang, Lian-Tao and Zhang, Cen",
    title = "{Unambiguously testing positivity at lepton colliders}",
    eprint = "2011.03055",
    archivePrefix = "arXiv",
    primaryClass = "hep-ph",
    reportNumber = "MITP/20-063",
    doi = "10.1103/PhysRevLett.129.011805",
    journal = "Phys. Rev. Lett.",
    volume = "129",
    number = "1",
    pages = "011805",
    year = "2022"
}

@article{Liu:2016idz,
    author = "Liu, Da and Pomarol, Alex and Rattazzi, Riccardo and Riva, Francesco",
    title = "{Patterns of Strong Coupling for LHC Searches}",
    eprint = "1603.03064",
    archivePrefix = "arXiv",
    primaryClass = "hep-ph",
    reportNumber = "CERN-TH-2016-048",
    doi = "10.1007/JHEP11(2016)141",
    journal = "JHEP",
    volume = "11",
    pages = "141",
    year = "2016"
}

@article{Jenkins:2013sda,
    author = "Jenkins, Elizabeth E. and Manohar, Aneesh V. and Trott, Michael",
    title = "{Naive dimensional analysis counting of gauge theory amplitudes and anomalous dimensions}",
    eprint = "1309.0819",
    archivePrefix = "arXiv",
    primaryClass = "hep-ph",
    reportNumber = "CERN-PH-TH-2013-213",
    doi = "10.1016/j.physletb.2013.09.020",
    journal = "Phys. Lett. B",
    volume = "726",
    pages = "697--702",
    year = "2013"
}

@article{Gasser:1983yg,
    author = "Gasser, J. and Leutwyler, H.",
    title = "{Chiral perturbation theory to one loop}",
    reportNumber = "CERN-TH-3689",
    doi = "10.1016/0003-4916(84)90242-2",
    journal = "Annals Phys.",
    volume = "158",
    pages = "142",
    year = "1984"
}

@article{Arkani-Hamed:2021ajd,
    author = "Arkani-Hamed, Nima and Huang, Yu-tin and Liu, Jin-Yu and Remmen, Grant N.",
    title = "{Causality, unitarity, and the weak gravity conjecture}",
    eprint = "2109.13937",
    archivePrefix = "arXiv",
    primaryClass = "hep-th",
    doi = "10.1007/JHEP03(2022)083",
    journal = "JHEP",
    volume = "03",
    pages = "083",
    year = "2022"
}

@article{Aoude:2024xpx,
    author = "Aoude, Rafael and Elor, Gilly and Remmen, Grant N. and Sumensari, Olcyr",
    title = "{Positivity in Amplitudes and Quantum Entanglement}",
    eprint = "2402.16956",
    archivePrefix = "arXiv",
    primaryClass = "hep-th",
    doi = "10.1002/prop.70113",
    journal = "Fortsch. Phys.",
    volume = "74",
    pages = "e70113",
    year = "2026"
}

@article{DuasoPueyo:2024usw,
    author = "Duaso Pueyo, Carlos and Goodhew, Harry and McCulloch, Ciaran and Pajer, Enrico",
    title = "{Perturbative unitarity bounds from momentum-space entanglement}",
    eprint = "2410.23709",
    archivePrefix = "arXiv",
    primaryClass = "hep-th",
    doi = "10.1007/JHEP08(2025)047",
    journal = "JHEP",
    volume = "08",
    pages = "047",
    year = "2025"
}

@article{Low:2024hvn,
    author = "Low, Ian and Yin, Zhewei",
    title = "{Elastic cross section is entanglement entropy}",
    eprint = "2410.22414",
    archivePrefix = "arXiv",
    primaryClass = "hep-th",
    doi = "10.1103/PhysRevD.111.065027",
    journal = "Phys. Rev. D",
    volume = "111",
    number = "6",
    pages = "065027",
    year = "2025"
}

@article{Low:2024mrk,
    author = "Low, Ian and Yin, Zhewei",
    title = "{Area law for entanglement entropy in particle scattering}",
    eprint = "2405.08056",
    archivePrefix = "arXiv",
    primaryClass = "hep-th",
    doi = "10.1103/3yg7-r5s9",
    journal = "Phys. Rev. D",
    volume = "113",
    number = "6",
    pages = "065004",
    year = "2026"
}

@article{Alonso:2015fsp,
    author = "Alonso, Rodrigo and Jenkins, Elizabeth E. and Manohar, Aneesh V.",
    title = "{A geometric formulation of Higgs effective field theory: measuring the curvature of scalar field space}",
    eprint = "1511.00724",
    archivePrefix = "arXiv",
    primaryClass = "hep-ph",
    reportNumber = "CERN-PH-TH-2015-257",
    doi = "10.1016/j.physletb.2016.01.041",
    journal = "Phys. Lett. B",
    volume = "754",
    pages = "335--342",
    year = "2016"
}

@article{Li:2024ciy,
    author = "Li, Xu-Xiang and Lu, Xiaochuan and Zhang, Zhengkang",
    title = "{The geometric universal one-loop effective action}",
    eprint = "2411.04173",
    archivePrefix = "arXiv",
    primaryClass = "hep-ph",
    doi = "10.1007/JHEP08(2025)102",
    journal = "JHEP",
    volume = "08",
    pages = "102",
    year = "2025"
}

@article{Aigner:2025xyt,
    author = "Aigner, Patrick and Bellafronte, Luigi and Gendy, Emanuele and Haslehner, Dominik and Weiler, Andreas",
    title = "{Renormalising the field-space geometry}",
    eprint = "2503.09785",
    archivePrefix = "arXiv",
    primaryClass = "hep-th",
    doi = "10.1007/JHEP07(2025)167",
    journal = "JHEP",
    volume = "07",
    pages = "167",
    year = "2025"
}

@article{DasBakshi:2023htx,
    author = "Das Bakshi, Supratim and D\'\i{}az-Carmona, \'Alvaro",
    title = "{Renormalisation of SMEFT bosonic interactions up to dimension eight by LNV operators}",
    eprint = "2301.07151",
    archivePrefix = "arXiv",
    primaryClass = "hep-ph",
    doi = "10.1007/JHEP06(2023)123",
    journal = "JHEP",
    volume = "06",
    pages = "123",
    year = "2023"
}

@article{Assi:2025fsm,
    author = "Assi, Beno{\^\i}t and Helset, Andreas and Pag{\`e}s, Julie and Shen, Chia-Hsien",
    title = "{Renormalizing two-fermion operators in the SMEFT via supergeometry}",
    eprint = "2504.18537",
    archivePrefix = "arXiv",
    primaryClass = "hep-ph",
    reportNumber = "CERN-TH-2025-084, FERMILAB-PUB-25-0280-V",
    doi = "10.1007/JHEP12(2025)082",
    journal = "JHEP",
    volume = "12",
    pages = "082",
    year = "2025"
}

@article{Weinberg:1966kf,
    author = "Weinberg, Steven",
    title = "{Pion scattering lengths}",
    doi = "10.1103/PhysRevLett.17.616",
    journal = "Phys. Rev. Lett.",
    volume = "17",
    pages = "616--621",
    year = "1966"
}

@article{Distler:2006if,
    author = "Distler, Jacques and Grinstein, Benjamin and Porto, Rafael A. and Rothstein, Ira Z.",
    title = "{Falsifying Models of New Physics via WW Scattering}",
    eprint = "hep-ph/0604255",
    archivePrefix = "arXiv",
    reportNumber = "UCSD-PTH-06-06, UTTTG-06-06, CMUHEP-06-07",
    doi = "10.1103/PhysRevLett.98.041601",
    journal = "Phys. Rev. Lett.",
    volume = "98",
    pages = "041601",
    year = "2007"
}

@article{Pennington:1994kc,
    author = "Pennington, M. R. and Portoles, J.",
    title = "{The Chiral Lagrangian parameters, l1, l2, are determined by the rho resonance}",
    eprint = "hep-ph/9409426",
    archivePrefix = "arXiv",
    reportNumber = "DTP-94-54",
    doi = "10.1016/0370-2693(94)01551-M",
    journal = "Phys. Lett. B",
    volume = "344",
    pages = "399--406",
    year = "1995"
}

@article{Ananthanarayan:1994hf,
    author = "Ananthanarayan, B. and Toublan, D. and Wanders, G.",
    title = "{Consistency of the chiral pion pion scattering amplitudes with axiomatic constraints}",
    eprint = "hep-ph/9410302",
    archivePrefix = "arXiv",
    reportNumber = "UNIL-TP-4-94",
    doi = "10.1103/PhysRevD.51.1093",
    journal = "Phys. Rev. D",
    volume = "51",
    pages = "1093--1100",
    year = "1995"
}

@article{Dita:1998mh,
    author = "Dita, Petre",
    title = "{Positivity constraints on chiral perturbation theory pion pion scattering amplitudes}",
    eprint = "hep-ph/9809568",
    archivePrefix = "arXiv",
    doi = "10.1103/PhysRevD.59.094007",
    journal = "Phys. Rev. D",
    volume = "59",
    pages = "094007",
    year = "1999"
}

@article{Cheung:2025nhw,
    author = "Cheung, Clifford and Remmen, Grant N.",
    title = "{Multipositivity bounds for scattering amplitudes}",
    eprint = "2505.05553",
    archivePrefix = "arXiv",
    primaryClass = "hep-th",
    reportNumber = "CALT-TH 2025-010",
    doi = "10.1103/wt4x-2149",
    journal = "Phys. Rev. D",
    volume = "112",
    number = "1",
    pages = "016017",
    year = "2025"
}

@article{Chandrasekaran:2018qmx,
    author = "Chandrasekaran, Venkatesa and Remmen, Grant N. and Shahbazi-Moghaddam, Arvin",
    title = "{Higher-Point Positivity}",
    eprint = "1804.03153",
    archivePrefix = "arXiv",
    primaryClass = "hep-th",
    doi = "10.1007/JHEP11(2018)015",
    journal = "JHEP",
    volume = "11",
    pages = "015",
    year = "2018"
}

@article{Gasser:1984gg,
    author = "Gasser, J. and Leutwyler, H.",
    title = "{Chiral Perturbation Theory: Expansions in the Mass of the Strange Quark}",
    reportNumber = "CERN-TH-3798",
    doi = "10.1016/0550-3213(85)90492-4",
    journal = "Nucl. Phys. B",
    volume = "250",
    pages = "465--516",
    year = "1985"
}

@article{deVries:2019nsu,
    author = "de Vries, Jordy and Falcioni, Giulio and Herzog, Franz and Ruijl, Ben",
    title = "{Two- and three-loop anomalous dimensions of Weinberg{\textquoteright}s dimension-six CP-odd gluonic operator}",
    eprint = "1907.04923",
    archivePrefix = "arXiv",
    primaryClass = "hep-ph",
    reportNumber = "RBRC-1318, Nikhef 2019-033",
    doi = "10.1103/PhysRevD.102.016010",
    journal = "Phys. Rev. D",
    volume = "102",
    number = "1",
    pages = "016010",
    year = "2020"
}

@article{Born:2024mgz,
    author = "Born, Lukas and Fuentes-Mart{\'\i}n, Javier and Kvedarait{\.{e}}, Sandra and Thomsen, Anders Eller",
    title = "{Two-loop running in the bosonic SMEFT using functional methods}",
    eprint = "2410.07320",
    archivePrefix = "arXiv",
    primaryClass = "hep-ph",
    doi = "10.1007/JHEP05(2025)121",
    journal = "JHEP",
    volume = "05",
    pages = "121",
    year = "2025"
}

@article{DiNoi:2024ajj,
    author = {Di Noi, Stefano and Gr{\"o}ber, Ramona and Mandal, Manoj K.},
    title = "{Two-loop running effects in Higgs physics in Standard Model Effective Field Theory}",
    eprint = "2408.03252",
    archivePrefix = "arXiv",
    primaryClass = "hep-ph",
    reportNumber = "COMETA-2024-19",
    doi = "10.1007/JHEP12(2024)220",
    journal = "JHEP",
    volume = "12",
    pages = "220",
    year = "2024"
}

@article{Catani:1996vz,
    author = "Catani, S. and Seymour, M. H.",
    title = "{A General algorithm for calculating jet cross-sections in NLO QCD}",
    eprint = "hep-ph/9605323",
    archivePrefix = "arXiv",
    reportNumber = "CERN-TH-96-029, CERN-TH-96-29",
    doi = "10.1016/S0550-3213(96)00589-5",
    journal = "Nucl. Phys. B",
    volume = "485",
    pages = "291--419",
    year = "1997",
    note = "[Erratum: Nucl.Phys.B 510, 503--504 (1998)]"
}

@article{Chakraborty:2024ciu,
    author = "Chakraborty, Debsubhra and Chattopadhyay, Susobhan and Gupta, Rick S.",
    title = "{Complete set of positivity constraints on the HEFT at NLO}",
    eprint = "2412.14155",
    archivePrefix = "arXiv",
    primaryClass = "hep-ph",
    doi = "10.1103/7qgz-9ykc",
    journal = "Phys. Rev. D",
    volume = "113",
    number = "5",
    pages = "053007",
    year = "2026"
}

@article{Aebischer:2025zxg,
    author = "Aebischer, Jason and Bresciani, Luigi C. and Selimovic, Nudzeim",
    title = "{Anomalous dimension of a general effective gauge theory. Part I. Bosonic sector}",
    eprint = "2502.14030",
    archivePrefix = "arXiv",
    primaryClass = "hep-ph",
    reportNumber = "CERN-TH-2025-032",
    doi = "10.1007/JHEP08(2025)209",
    journal = "JHEP",
    volume = "08",
    pages = "209",
    year = "2025"
}

@misc{Supp,
  note = {The supplemental material is available after references and it contains a reference to~\cite{Catani:1996vz}}
}

\newpage
\phantom{.}
\newpage 

\widetext
\begin{center}
\textbf{\large Supplemental Material}
\end{center}
\vspace{-0.2cm}
\setcounter{equation}{0}
\setcounter{figure}{0}
\setcounter{table}{0}
\setcounter{page}{1}
\makeatletter
\renewcommand{\theequation}{S\arabic{equation}}
\renewcommand{\thefigure}{S\arabic{figure}}
\renewcommand{\thepage}{S\arabic{page}}

\sectionskip
\section{Examples of RG equations with definite signs}
We demonstrate the predictions of our main theorem
with examples in the SMEFT and the EFT of gravity coupled to a single real scalar. In the SMEFT, we consider the full RG of the $H^4D^4$ operators induced by a pair of dimension\nobreakdash-6 operators.
This expands on the example provided in \cite{Chala:2021wpj} with operators that cannot be generated at tree level by matching to weakly coupled UV completions. In particular, we also include contributions from $F^3$-type operators, which are computed here for the first time. 
In addition, we show that 
our theorem can constrain the RG of dimension-6 operators in the SMEFT, although this requires a more careful inspection of the channels that contribute.

We define $\dot c_i \equiv 16\pi^2 \tfrac{d\,c_i}{d \ln \mu}$ throughout the supplemental material.
We use the EFT Lagrangian in the form
\begin{align}
    \cL = \sum_{i,n} \frac{c_{i}}{\Lambda^{n-4}} \mathcal{O}_{n,i}\,,
\end{align}
where we sum over dimension-$n$ operators 
$\mathcal{O}_{n,i}$ with the Wilson coefficient as $c_i$, and $\Lambda$ is the UV cutoff. 
The operators will be defined below for different EFTs.

\subsection{The RG of $H^4D^4$ operators in the SMEFT}
We start with the renormalization of the $H^4D^4$ operators in the SMEFT at dimension eight, induced by pairs of dimension-6 insertions. The relevant operators are defined in Tab.~\ref{tab:opreators}, following \cite{Grzadkowski:2010es, Murphy:2020rsh}. For the clarity of the presentation, we define the following combinations $\mathcal{G}_{i}$ for operators with gauge bosons
\begin{align}
\mathcal{G}_{1}&\equiv
6\,g_{2}^{2}\left(c_{\W}^2+c_{\Wd}^2\right) ,\\[2pt]
\mathcal{G}_{2}&\equiv8\left(c_{\HWB}^2+c_{\HWBd}^2\right) ,\\[2pt]
\mathcal{G}_{3}&\equiv16\left(
c_{\HB}^2+c_{\HBd}^2+3\left( c_{\HW}^2+c_{\HWd}^2\right)+8\left(c_{\HG}^2+c_{\HGd}^2\right)\right) ,
\end{align}
and quantities $\mathcal{F}_{i}$ for operators with fermions
\begin{align}
\mathcal{F}_{1}&\equiv\frac{8}{3}\left[
3\,\tr{(c_{Hd})^{\dagger}c_{Hd}}+\tr{(c_{He})^{\dagger}c_{He}}+2\,\tr{(c_{HL}^{(1)})^{\dagger}c_{HL}^{(1)}}+6\,\tr{(c_{Hq}^{(1)})^{\dagger}c_{Hq}^{(1)}}+3\,\tr{(c_{Hu})^{\dagger}c_{Hu}}\right],\\[2pt]
\mathcal{F}_{2}&\equiv\frac{16}{3}\left[\tr{(c_{HL}^{(3)})^{\dagger}c_{HL}^{(3)}}+3\,\tr{(c_{Hq}^{(3)})^{\dagger}c_{Hq}^{(3)}}\right],\\[2pt]
\mathcal{F}_{3}&\equiv
8\,\tr{(c_{Hud})^{\dagger}c_{Hud}},
\end{align}
where the trace sums over the fermion flavor indices. Crucially, all combinations are positive definite
\begin{align}
    \mathcal{G}_{i}\ge 0\,,\qquad \mathcal{F}_{i} \ge 0\,.
\end{align}
The RG equations for $H^4D^4$ operators from inserting two dimension-6 operators are
\begin{align}
     \dot{c}_{H^4D^4}^{(1)} &= \frac13\left(
     -16 \,c_{H\square}^2
    +32\,c_{HD}c_{H\square}
    -11\,c_{HD}^2\right)-\left(
    \mathcal{G}_{1}+2\,\mathcal{G}_{2}\right)
    +\left(\mathcal{F}_{1}-\mathcal{F}_{2}-\mathcal{F}_{3}\right)
    ,\\[2mm]
     \dot{c}_{H^4D^4}^{(2)} & =  
    -\frac13 \left(
    16\,c_{H\square}^2+
    16\,c_{HD}c_{H\square}
    +5\,c_{HD}^2\right)
    -\mathcal{G}_{1} 
    -\left(\mathcal{F}_{1}+\mathcal{F}_{2}\right)
    ,\\[2mm]
     \dot{c}_{H^4D^4}^{(3)} &=
     \frac13\left(
     -40\, c_{H\square}^2
    -16\,c_{HD}c_{H\square}
    +7\,c_{HD}^2\right)
    +\left(2\,\mathcal{G}_{1}
    +\mathcal{G}_{2}
    -\mathcal{G}_{3}\right)
    +\left(2\,\mathcal{F}_{2}+\mathcal{F}_{3}\right)
    .
\end{align}
These results were previously computed in~\cite{Chala:2021pll,Helset:2022pde} for all operators except for the new contributions from the $F^3$, $\tilde F F^2$  and $H^2 \tilde{F}F$ operators.

Changing basis to the operators that can be probed in various forward limits~\cite{Remmen:2019cyz},
the RG equations become
\begin{align}
\dot{c}_{H^4D^4}^{(2)}&= 
-\frac{1}{3}\left(4 \,          (2\,c_{H\square}
+c_{HD})^2
    +c_{HD}^2\right)
    -\mathcal{G}_{1}
    -\left(\mathcal{F}_{1}+\mathcal{F}_{2}\right)
    ,\\[2mm]
\dot{c}_{H^4D^4}^{(1)}
+\dot{c}_{H^4D^4}^{(2)}
&=-\frac13
\left(28\,c_{H\square}^{2}
+4\,(c_{H\square}-2\,c_{HD})^{2}
\right)
-\left(2\,\mathcal{G}_{1}+2\,\mathcal{G}_{2}\right)
-\left(2\,\mathcal{F}_{2}+\mathcal{F}_{3} \right)
    ,\\[2mm]
\dot{c}_{H^4D^4}^{(1)}+\dot{c}_{H^4D^4}^{(2)}+\dot{c}_{H^4D^4}^{(3)}&=
-\left(24\,c_{H\square}^{2}+3\,c_{HD}^{2}\right)
-\left(\mathcal{G}_{2}
+\mathcal{G}_{3}\right),
\end{align}
which are manifestly sign-definite, as expected.
In addition, we remark that double insertions of fermionic operators do not contribute to
$\dot{c}_{H^4D^4}^{(1)}+\dot{c}_{H^4D^4}^{(2)}+\dot{c}_{H^4D^4}^{(3)}$.
This can be explained using the real scalar field parametrization of the Higgs doublet, $H=(\phi_1 +i \phi_2, \phi_3 +i\phi_4)^T/\sqrt{2}$, noting that the Feynman rule of the four-point vertex with the same real-scalar, $\phi_i^4$, is proportional to ${c}_{H^4D^4}^{(1)}+{c}_{H^4D^4}^{(2)}+{c}_{H^4D^4}^{(3)}$. The fermionic operators do not renormalize this vertex because they couple to the scalars through $\phi_i \partial_\mu \phi_j - \left(\partial_\mu \phi_i \right) \phi_j$.

\begin{table}[t]
\renewcommand{\arraystretch}{1.8}
\setlength{\tabcolsep}{4pt}

\begin{center}
\begin{minipage}[t]{2cm}

\begin{align*}
\begin{tabular}{c|c}
\multicolumn{2}{c}{$\bm{H^2F^2}$}\\
\hline
$\Op_{\HB}$&
$(H^{\dagger}H)B_{\mu\nu}B^{\mu\nu}$\\
$\Op_{\HBd}$&
$(H^{\dagger}H)\widetilde{B}_{\mu\nu}B^{\mu\nu}$\\
$\Op_{\HW}$&
$(H^{\dagger}H)W_{\mu\nu}^{I}W^{I\mu\nu}$\\
$\Op_{\HWd}$&
$(H^{\dagger}H)\widetilde{B}_{\mu\nu}^{I}W^{I\mu\nu}$\\
$\Op_{\HG}$&
$(H^{\dagger}H)G_{\mu\nu}^{A}G^{A\mu\nu}$\\
$\Op_{\HGd}$&
$(H^{\dagger}H)\widetilde{G}_{\mu\nu}^{A}G^{A\mu\nu}$\\
$\Op_{\HWB}$&
$(H^{\dagger}\tau^{I}H)W_{\mu\nu}^{I}B^{\mu\nu}$\\
$\Op_{\HWBd}$&
$(H^{\dagger}\tau^{I}H)\widetilde{W}_{\mu\nu}^{I}B^{\mu\nu}$\\
\multicolumn{2}{c}{}\\
\multicolumn{2}{c}{$\bm{F^3}$}\\
\hline
$\Op_{\W}$&
$\epsilon^{IJK}W_{\mu}^{I\nu}W_{\nu}^{J\rho}W_{\rho}^{K\mu}$\\
$\Op_{\Wd}$&
$\epsilon^{IJK}\widetilde{W}_{\mu}^{I\nu}W_{\nu}^{J\rho}W_{\rho}^{K\mu}$
\end{tabular}
\end{align*}
\end{minipage}
\hspace{1cm}
\begin{minipage}[t]{2cm}

\begin{align*}
\begin{tabular}{c|c}
\multicolumn{2}{c}{$\bm{\overline{\psi}\psi H^2D}$}\\
\hline
$\Op_{H L}^{(1)}$&
$(H^{\dagger}i \overset\leftrightarrow{D_{\mu}}H)(\overline{L}_{\alpha}\gamma^{\mu}L_{\beta})$\\
$\Op_{H L}^{(3)}$&
$(H^{\dagger}i \overset\leftrightarrow{D_{\mu}^{I}}H)(\overline{L}_{\alpha}\tau^{I}\gamma^{\mu}L_{\beta})$\\
$\Op_{H q}^{(1)}$&
$(H^{\dagger}i \overset\leftrightarrow{D_{\mu}}H)(\overline{q}_{\alpha}\gamma^{\mu}q_{\beta})$\\
$\Op_{H q}^{(3)}$&
$(H^{\dagger}i \overset\leftrightarrow{D_{\mu}^{I}}H)(\overline{q}_{\alpha}\tau^{I}\gamma^{\mu}q_{\beta})$\\
$\Op_{H e}$&
$(H^{\dagger}i \overset\leftrightarrow{D_{\mu}}H)(\overline{e}_{\alpha}\gamma^{\mu}e_{\beta})$\\
$\Op_{H u}$&
$(H^{\dagger}i \overset\leftrightarrow{D_{\mu}}H)(\overline{u}_{\alpha}\gamma^{\mu}u_{\beta})$\\
$\Op_{H d}$&
$(H^{\dagger}i \overset\leftrightarrow{D_{\mu}}H)(\overline{d}_{\alpha}\gamma^{\mu}d_{\beta})$\\
 $\Op_{Hud}$&
$i(\widetilde{H}^{\dagger} D_{\mu}H(\overline{u}_{\alpha}\gamma^{\mu}d_{\beta})$\\
\multicolumn{2}{c}{}\\
\multicolumn{2}{c}{$\bm{\left(\overline{\psi}\psi\right)\left(\overline{\psi}\psi\right)}$}\\
\hline
$\Op_{LL}$
&
$\left(\overline{L}_\alpha\gamma_{\mu}L_\beta\right)\left(\overline{L}_\gamma \gamma^{\mu}L_\delta \right)$ 
\end{tabular}
\end{align*}

\end{minipage}
\hspace{1cm}
\begin{minipage}[t]{2cm}

\begin{align*}
\begin{tabular}{c|c}
\multicolumn{2}{c}{$\bm{H^4D^2}$}\\
\hline
$\Op_{\HD}$&
$(H^{\dagger}D_{\mu}H)^{*}(H^{\dagger}D^{\mu}H)$\\
$\Op_{\Hb}$&
$(H^{\dagger}H)\square(H^{\dagger}H)$\\
\multicolumn{2}{c}{}\\
\multicolumn{2}{c}{$\bm{H^4D^4}$}\\
\hline
$\Op_{H^{4}D^{4}}^{(1)}$&
$(D_\mu H^\dagger D_\nu H)(D^\nu H^\dagger D^\mu H)$\\
$\Op_{H^4D^4}^{(2)}$ &
$(D_\mu H^\dagger D_\nu H)(D^\mu H^\dagger D^\nu H)$\\
$\Op_{H^4D^4}^{(3)}$&
$(D_\mu H^\dagger D^\mu H)(D_\nu H^\dagger D^\nu H)$
\end{tabular}
\end{align*}

\end{minipage}
\end{center}
\caption{SMEFT operators at dimension six and eight used in this work, following the basis of~\cite{Grzadkowski:2010es,Murphy:2020rsh}.}
\label{tab:opreators}
\end{table}


\subsection{The RG of dimension-6 operators in the SMEFT}

In the main text,
we showed that the $s$ and $u$ channel cut diagrams are positive in the FL (for positive $s$ and $u$). However, since $u=-s$, this generally does not lead to a definite sign in the RG of dimension-6 operators. Nevertheless, in the main text, we showed that the mixing of two insertions of the dimension-5 Weinberg operator~\cite{Weinberg:1979sa} into the $H^4D^2$ operators is definite in sign.
We provide more details on this calculation and we show that the mixing of the Weinberg operator into the two-lepton and four-lepton operators is also sign-definite.
In all cases, the sign of the RG can be determined
because the $u$ channel is manifestly absent.

\begin{figure}[t]
  \centering
    \includegraphics[page=3,trim={0 28.4cm 53cm 0},clip,scale=0.23]{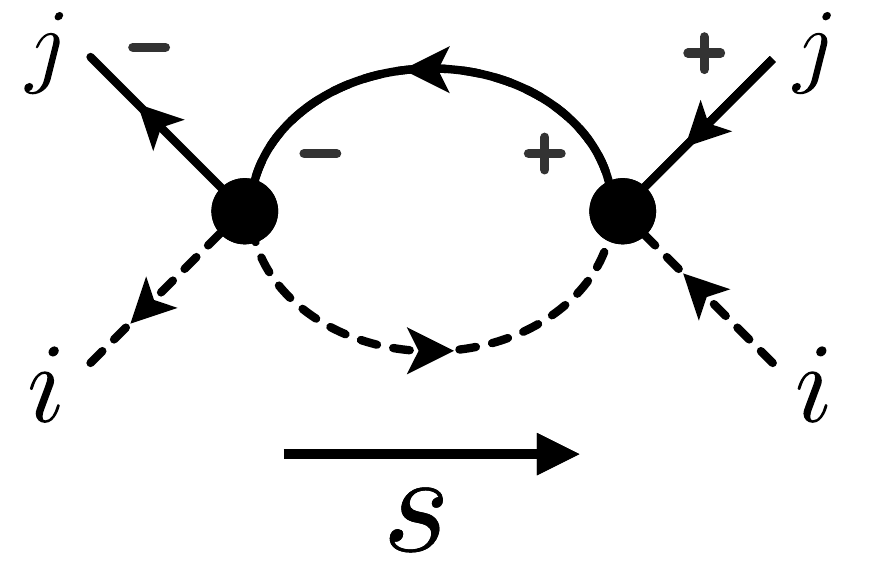}
  \hspace{0.06\textwidth}
  \raisebox{12mm}{;}
  \hspace{0.08\textwidth}  
    \includegraphics[page=4,trim={0 28.4cm 53.4cm 0},clip,scale=0.23]{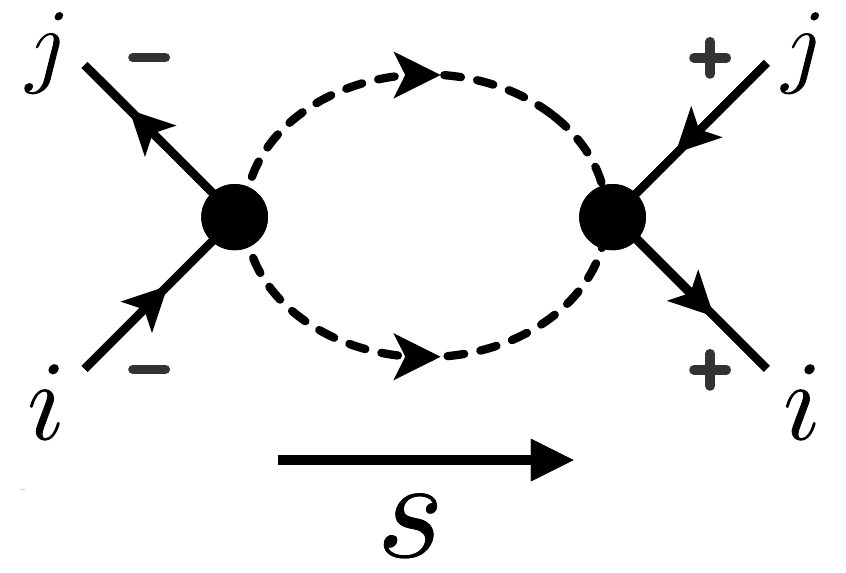}
  \caption{Diagrams for the mixing of the dimension-5 Weinberg operator into $\mathcal{O}_{HL}^{(1)}$ and $\mathcal{O}^{(3)}_{HL}$ (left) and into $\mathcal{O}_{LL}$ (right). Dashed lines and solid lines correspond to Higgs boson and lepton doublets, respectively. Helicities are indicated in the all outgoing convention.
  The $u$ channel is omitted for the mixing into $\mathcal{O}_{HL}^{(1,3)}$ because it contributes to the opposite choice of helicities. The same holds for the $t$-channel diagram for the mixing into 
  $\mathcal{O}_{LL}$.}
  \label{fig:weinberg}
\end{figure}

We define the dimension-5 SMEFT as
\begin{align}
\cL_5 = \frac{c^{\alpha \beta}_{5}}{2\Lambda}  \left(\overline{L}{}^{i}_\alpha L^c
{}_\beta^k \right)
\epsilon_{ij} \epsilon_{kl} H^{\dagger \,j} H^{\dagger \,l} +
\frac{c^{\alpha \beta*}_{5}}{2\Lambda}  \left(\overline{L^c} 
{}_\beta^k L_\alpha^i \right)
\epsilon_{ij} \epsilon_{kl} H^{j} H^{l}\,,
\end{align}
where $L^c = C \, \overline{L}^T$ and $C$ is the charge conjugation matrix that numerically equals to $i\gamma^0\gamma^2$.
We use Roman letters for $SU(2)$ indices and Greek letters for flavor indices, and $\epsilon$ is the anti-symmetric tensor with $\epsilon_{12}=1$.
The coupling $c^{\alpha \beta}_5=c^{\beta\alpha }_5$ since $\overline{L}{}^{i}_\alpha L^c{}_\beta^k=\overline{L}{}^{k}_\beta L^c{}_\alpha^i$.
The Weinberg operator leads to the four-point amplitudes
\begin{align}
{A}(H_i^\dagger(p_1) L_{j\alpha}^{c+}(p_2)
    \to L^-_{k\beta}(p_3)H_l(p_4))
    &= c_5^{\alpha\beta}
    \left(2\,\delta_{il}\delta_{jk} -  \delta_{ij}\delta_{kl}
    - \delta_{ik}\delta_{jl}\right) \langle 32 \rangle \,,\\
{A}(H_i(p_1) L_{j\alpha}^{-}(p_2)
    \to L^{c+}_{k\beta}(p_3)H^\dagger_l(p_4))
    &= c_5^{*\alpha\beta}
    \left(2\,\delta_{il}\delta_{jk} -  \delta_{ij}\delta_{kl}
    - \delta_{ik}\delta_{jl}\right) [ 23 ] \,.
\end{align}
In the arguments of $A(i\rightarrow f)$, the helicities of the initial and final states are labeled in the incoming and outgoing conventions which differs from the all-outgoing convention used in the figures.

Four-point diagrams with two insertions of the Weinberg operators generate the $H^4D^2$, $\overline{L}L H^2D$, and $\overline{L}L \overline{L}L$ operators, which are defined in Tab.~\ref{tab:opreators}. All their RG equations exhibit sign-definite behavior.
The $H^4D^2$ operators have been discussed in the main text. 
Secondly, the two-lepton operators generate the forward amplitude
\begin{equation}
    {A}(H_i(p_1) L_{j\beta}^-(p_2)
    \to L^-_{j\alpha}(p_2)H_i(p_1)) 
    = 2 \left( c_{H L}^{(1)\alpha\beta} \delta_{ii} \delta_{jj}
            + c_{H L}^{(3)\alpha\beta} (
            2\,\delta_{ij} \delta_{ij} - 
            \delta_{ii} \delta_{jj})
        \right) 
      s  \,.  
\end{equation}
where $c_{HL}^{(1)}$ and $c_{HL}^{(3)}$ are Hermitian matrices.
The $i,j$ indices above are not summed.
Also in this case, 
only the $s$ channel is non-zero for the RG contribution from insertions of the Weinberg operators, see Fig.~\ref{fig:weinberg}. The $u$-channel contributes when the opposite helicities are chosen.
This leads to the prediction on the RG,
\begin{align}
    &\dot c_{HL}^{(1)} -  \dot c_{HL}^{(3)} \le 0 \,, \qquad
    \dot c_{HL}^{(1)} +  \dot c_{HL}^{(3)} \le 0 \,.
\end{align}
Since we have not restricted to the diagonal element in flavor space, the above inequality should be viewed as the negative definite on the matrix, $\langle v| M |v \rangle \le 0$ for any vector $|v\rangle$.
Indeed, we find that the explicit results~\cite{Davidson:2018zuo}
\begin{align}
    \dot c_{H L}^{(1)\alpha\beta}
    &= -\frac{3}{2} 
    (c_5 c_5^{*})^{\alpha\beta} \,, \qquad
    \dot c_{H L}^{(3)\alpha\beta}
    =  
    \left(c_5 c_5^*\right)^{\alpha \beta} \,,
\end{align}
are consistent with our theorem.
Here the $c_{H L}^{(i)\alpha\beta}$ is the Wilson coefficient for $\Op_{H L}^{(1)}$ in Tab.~\ref{tab:opreators}.

Finally, the relevant four-lepton operator, $\mathcal{O}_{LL}$, with Wilson coefficient $c_{LL}^{\alpha\beta\gamma\delta}$ in Tab.~\ref{tab:opreators}, leads to the forward amplitude
\begin{align}
    {A}(L_{i\beta}^-(p_1) L_{j\delta}^-(p_2) \to L_{j\gamma}^-(p_2) L_{i\alpha}^-(p_1)) &= 4 \, s \left(c_{LL}^{\alpha\beta\gamma\delta}\delta_{ii}\delta_{jj}
    +c_{LL}^{\alpha\delta\gamma\beta} \delta_{ij} \delta_{ij}\right),
\end{align}
where $i,j$ are not summed.
In this case, only the $s$-channel diagram exists, see Fig.~\ref{fig:weinberg}. 
Since this channel contributes with definite sign in the FL, the forward elements of the tensor have negative RG, $\langle v \otimes w |\dot c_{LL}| w\otimes v\rangle \le 0$, 
where $v$ and $w$ are the flavor wavefunctions of particle $1$ and $2$.
This is confirmed by the explicit RG~\cite{Davidson:2018zuo},
\begin{equation}
    \dot{c}_{LL}^{\alpha\beta\gamma\delta} = -\frac{1}{2} c_5^{\alpha \gamma} c_5^{*\beta\delta}.
\end{equation}
Therefore, 
our theorem determines the sign of all the RG effects on dimension-6 operators arising from inserting two Weinberg operators.

\subsection{Gravitational EFT}

Our derivation of the signs in the RG
also applies in the context of gravity. We consider a single scalar field coupled to gravity with interaction $R^2\phi^2$. We define this operator through the on-shell amplitude~\cite{Baratella:2021guc}
\begin{equation}\label{14}
    A_{R^{2}\phi^2}(1_{\phi},2_{\phi},3_{h_{+}},4_{h_{+}})=\frac{c_{R^{2}\phi^{2}}}{M_{Pl}^{2}
    }
    [34]^{4}\,.
\end{equation}

\begin{figure}[t]
\begin{center}
\includegraphics[page=5,
trim={0 28.4cm 53cm 0},
clip,scale=0.30]{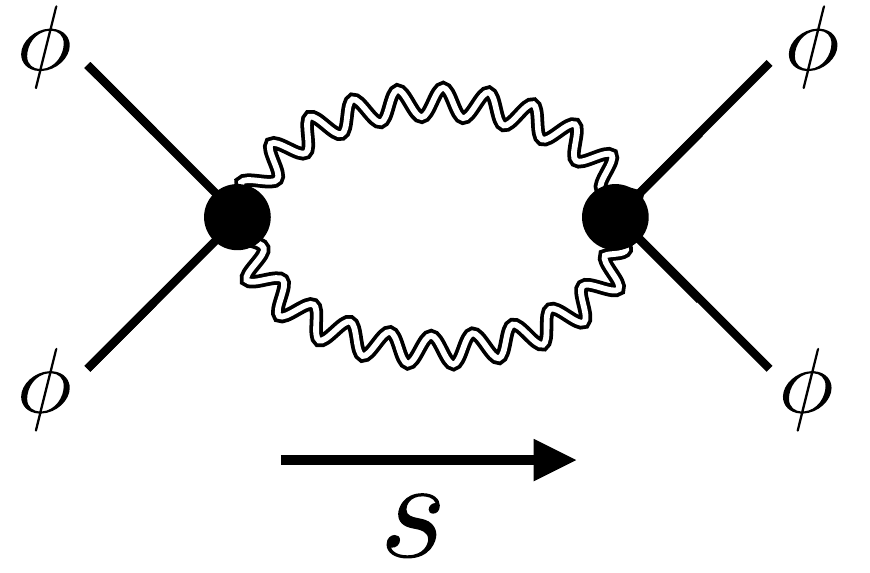}
\raisebox{16.5mm}{\Large $\phantom{.}+ (s\to t) + (s\to u)$}
\end{center}
\vspace{-0.5cm}
\caption{\label{fig:Gravity} 
Diagrams for the mixing of two insertions of $R^2\phi^2$ into $\phi^4\partial^8$.
}
\end{figure}

With double insertions, this operator mixes into the four-scalar operator at dimension twelve, $\phi^4\partial^8$, see Fig.~\ref{fig:Gravity}. We define this operator by
\begin{equation}
    A_{\phi^4\partial^8}(1_{\phi},2_{\phi},3_{\phi},4_{\phi})
    =\frac{c_{\phi^4\partial^8}}{\Lambda^8}\left(s^2+t^2+u^2\right)^{2}\,.
\end{equation}
The RG of $c_{\phi^4\partial^8}$ is given by
\begin{equation}
\dot{c}_{\phi^4\partial^8}=
-c_{R^2\phi^2}^2
\,,
\end{equation}
where the sign is as expected from the theorem in the main text.

\sectionskip
\section{Violation of Tree-level Positivity from RG flow}
In this section, we discuss the potential violation of tree-level positivity bounds by RG running, in settings where our theorem does not apply.
In all examples, the RG equations have a positive sign, such that the coefficient is driven toward smaller values in the IR.
However, the violation of positivity is prohibited in the first example because $c=0$ is a fixed point.
We provide a second example, which is more similar to the SMEFT, that bypasses this obstruction and thereby results in a violation of tree-level positivity bounds.
See also~\cite{Chala:2021wpj} for a similar example. 
In the third example, we consider diagrams that include three-point vertices, and show that manifestly positive integrands may nevertheless lead to RG equations with a positive sign. Finally, we discuss the consistency of our findings with dispersive arguments for positivity bounds.

\subsection{Toy model UV completion}
\label{sec:negative_model1}
In this example we consider the full theory
\begin{align}
    \cL_{\rm full} &= \frac{1}{2} (\partial \phi)^2 
    +\frac{1}{2} (\partial \chi)^2 -\frac{1}{2} \Lambda^2 \chi^2
    -\frac{\lambda}{4!}\phi^4 -\frac{g}{4} \phi^2 \chi^2 \,,
\end{align}
and obtain the EFT of $\phi$ by integrating out the heavy scalar $\chi$\,,
\begin{align}
    \cL &=\frac{1}{2}(\partial\phi)^{2}-\frac{\lambda}{4!}\phi^{4}+\frac{c_{8}}{\Lambda^{4}}(\partial\phi)^{4}+\dots \,.
    \label{eq:single_phi_EFT}
\end{align}
We suppress the $\phi^6$ operator, because it does not mix into $(\partial\phi)^4$ at one loop.
We also assume that $g \ll \lambda$ so $\lambda$ does not get significant matching corrections.
This is the weakly-coupled EFT limit, in which 
we do not predict constrains on
the overall direction of the RG flow.
The EFT matching (see the first diagram in Fig.~\ref{fig:matching}) yields
\begin{align}
    c_8 = \frac{g^2}{16\pi^2} \frac{1}{240}\,,
\end{align}
which agrees with the positivity bound, $c_8>0$~\cite{Adams:2006sv}.
Moreover, the RG in the IR is
\begin{align}
    \dot{c}_8 &= \frac{10}{3}\lambda \,c_8 \,,
\end{align}
which drives $c_8$ toward smaller values in the IR.
However, its value never crosses zero since $c_8=0$ is a fixed point of the RG (at one loop).
To demonstrate a violation of the tree-level positivity bounds, we must consider an example with multiple dimension-8 four-point operators, as is the case in the Higgs sector of the SMEFT.

\subsection{Extended Higgs Sector}
\label{sec:negative_model2}
\begin{figure}[t]
  \centering
    \includegraphics[page=7,trim={0 27.1cm 53cm 0},clip,scale=0.27]{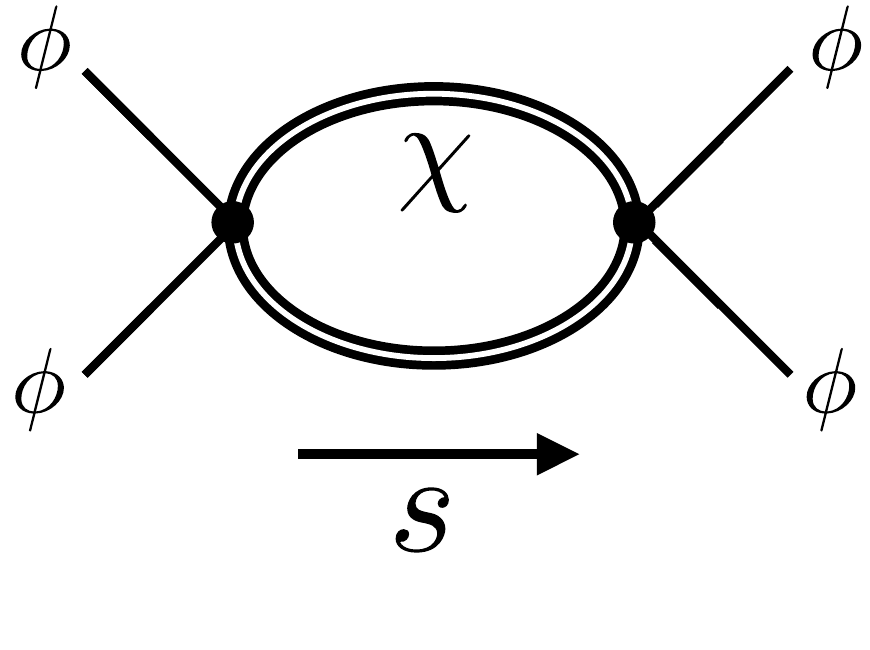}
  \raisebox{1.8cm}{\Large $\phantom{.}+ (s\to t) + (s\to u)$ \quad;}
  \hspace{0.02\textwidth}
    \includegraphics[page=8,trim={0 27.1cm 52.3cm 0},clip,scale=0.27]{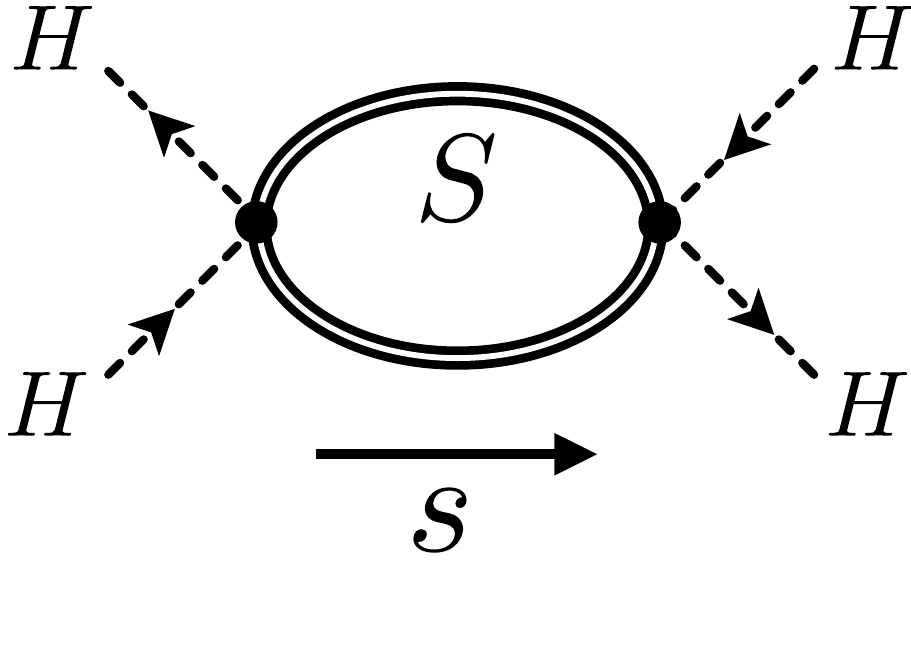}
\raisebox{1.8cm}{\hspace{-1mm} \Large $\phantom{.}+ (s\to t)$}
\raisebox{3mm}{
\hspace{6mm}\textbf{
Toy model UV completion}
  \hspace{0.27\textwidth}
\textbf{
Extended Higgs sector}
}
    
  \caption{Diagrams for the matching calculation in the toy model UV completion (left) and in the extended Higgs sector (right). The double lines denote the heavy particles $\chi$ and $S$ in the left and right diagrams, respectively.}
\label{fig:matching}
\end{figure}

We consider a UV completion of the Higgs sector given by
\begin{align}
    \cL_{\rm full} &= 
    \frac{1}{2} (\partial S)^2 -\frac{1}{2} \Lambda^2 S^2
    +\partial H^\dagger \partial H -m^2 H^\dagger H-\lambda\, (H^\dagger H)^2 -\frac{g}{2} S^2 H^\dagger H\,.
\end{align}
We consider $\lambda >0$ and the hierarchy of scales $m \ll \Lambda$. The coupling $g$ is set as $g \sim m^2/\Lambda^2$ to preserve the IR scale $m$.
This is similar to the examples considered in~\cite{Chala:2021wpj}.
The EFT after integrating out the singlet $S$ yields the scalar sector of the SMEFT. 
\begin{align}
    \cL &= 
    \partial H^\dagger \partial H
    -\lambda\, (H^\dagger H)^2
    +\frac{1}{\Lambda^2}\left(c_{HD}\, \Op_{HD} +c_{H\Box}\, \Op_{H\Box} \right)
    +\frac{1}{\Lambda^4}\sum^3_{i=1}\left(c^{(i)}_{H^4D^4}\, \Op^{(i)}_{H^4D^4}
    \right),
\end{align}
where the higher-dimensional operators are defined in Tab.~\ref{tab:opreators}.
To simplify the RG analysis, we have fine-tuned $g$ such that the Higgs has no mass in the IR.
The one-loop matching results from the right diagram of Fig.~\ref{fig:matching} are
\begin{align}
    \left(c_{HD}, \, c_{H\square}\right)
     &=\left(0, -\frac{1}{24} \frac{g^2}{16\pi^2} \right)\,,\\
    \left(c_{H^4D^4}^{(1)}, \, c_{H^4D^4}^{(2)},\, 
    c_{H^4D^4}^{(3)} \right) &= \left(0,0,\frac{1}{60} \frac{g^2}{16\pi^2} \right).
    \label{S33}
\end{align}
Because the heavy particle $S$ is an $SU(2)$ singlet, the $SU(2)$ indices of Higgs cannot contract across the other side of the loop in the right diagram of Fig.~\ref{fig:matching}. This leads to the zeros of $c_{HD}$, $c_{H^4D^4}^{(1)}$ and $c_{H^4D^4}^{(2)}$. (The same zeros also occur in \cite{Chala:2021wpj}.)

Since $g \ll \lambda$, the RG contributions to $c^{(i)}_{H^4D^4}$ from dimension-6 operators can be ignored compared to the interference between SM and dimension-8 operators. 
The RG contributions from the mass term interfere with beyond-dimension-8 operators are also suppressed because we have fine-tuned the mass to be negligible in the current energy scale.
The relevant RG equations are then~\cite{AccettulliHuber:2021uoa}
\newcommand{\cc}[2]{c_{#1,#2}}
\begin{align}
    \dot{c}_{H^{4}D^4}^{(2)} &=
    \frac{8}{3} \lambda \left(c_{H^4D^4}^{(1)}+3\,c_{H^4D^4}^{(2)}+c_{H^4D^4}^{(3)}\right)
    \hspace{6.3mm}>0 \,,\\
    \dot{c}_{H^4D^4}^{(1)}+\dot{c}_{H^4D^4}^{(2)} &=
    \frac{16}{3} \lambda \left(2\,c_{H^4D^4}^{(1)}+2\,c_{H^4D^4}^{(2)}+c_{H^4D^4}^{(3)}\right)
    \hspace{2.3mm}>0 \,,\\
    \dot{c}_{H^4D^4}^{(1)}+\dot{c}_{H^4D^4}^{(2)}+\dot{c}_{H^4D^4}^{(3)} &=
    \frac{16}{3} \lambda \left(5\,c_{H^4D^4}^{(1)}+4\,c_{H^4D^4}^{(2)}+6\,c_{H^4D^4}^{(3)}\right) >0 \,,
\end{align}
where we deduce the signs from the matching conditions \eqref{S33}.
Since $c_{H^4D^4}^{(2)}$ and $c_{H^4D^4}^{(1)}+c_{H^4D^4}^{(2)}$ are zero at the matching scale, the tree-level positivity bounds for both are violated when we run to $E<\Lambda$ using the RG equations,
\begin{align}
    c_{H^{4}D^4}^{(2)}(E) &= -\frac{2}{45}\frac{\lambda}{16\pi^2}\frac{g^2}{16\pi^2} \ln\!\left(\frac{\Lambda}{E} \right),
    \nn \\
    c_{H^{4}D^4}^{(1)}(E)+c_{H^{4}D^4}^{(2)}(E) &= -\frac{4}{45}\frac{\lambda}{16\pi^2}\frac{g^2}{16\pi^2} \ln\!\left(\frac{\Lambda}{E} \right).
    \label{eq:negative}
\end{align}
Because the operators are generated at the one-loop level at the matching scale, this RG effect is comparable to the two-loop matching results at the order $\order\!\left(\lambda \, g^2\right)$.
However, the RG effects in \eqref{eq:negative} are always dominant by the large logarithms when we run into deep IR, $E \ll \Lambda$ (but $E$ is still much larger than the mass). This establishes a violation of the tree-level positivity bounds in a unitary and causal UV model. Therefore, one should not directly use the tree-level positivity bounds as robust theoretical priors, as shown in \cite{Chala:2021wpj} already.

\subsection{\sug{Including three-point interactions}}\label{sec:3pt}
In our derivation of the signs in the RG, we crucially assume the absence of three-point interactions. Here we examine a concrete example with three-point interactions and show that one does not have control of the sign of the RG running, even if the amplitude is IR finite. This obstructs a naive extension of our theorem to cases with three-point interactions.

We consider the model of a complex  scalar $\chi$ charged under a $U(1)$ gauge symmetry and a neutral scalar $\phi$ with Lagrangian
\begin{align}
    \mathcal{L} = D_\mu\bar\chi \, D^\mu \chi
    +\frac12\partial_\mu\phi\partial^\mu\phi 
    -\frac14 F_{\mu\nu}F^{\mu\nu} 
    + \frac12 \frac{c_5}{\Lambda} \phi F^2
    + \frac12 \frac{c_6}{\Lambda^2} \phi^2D_\mu \bar\chi \, D^\mu \chi\,,
\end{align}
where we set the leading contact interactions between the scalar fields to zero for expository simplicity.  We are interested in the RG running of $c_6$ induced by two insertions of $c_5$. Even though $c_6$ is a dimension-6 Wilson coefficient, to which our theorem does not directly apply, this mixing will serve our purpose of illustrating that a positive integrand may lead to opposite-sign contributions to the RG when three-point interactions are present.

\begin{figure}
    \centering 
    \includegraphics[page=1,trim={13cm 27.1cm 40cm 0},clip,scale=0.32]{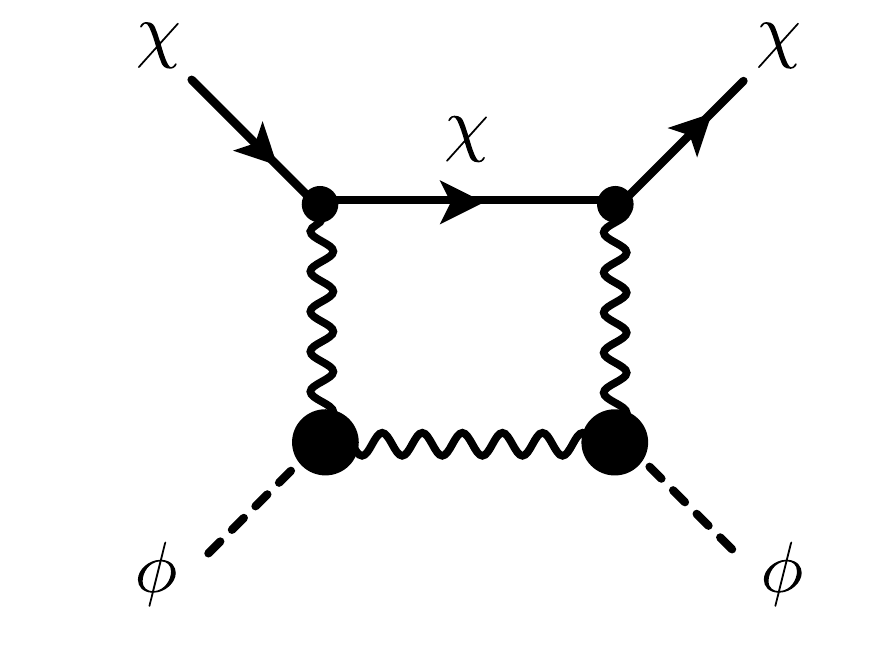} 
    \includegraphics[page=1,trim={0 27.1cm 53cm 0},clip,scale=0.32]{Diagrams3pt.pdf}\hspace{-1.1cm}
    \includegraphics[page=1,trim={29.3cm 27.1cm 24cm 0},clip,scale=0.32]{Diagrams3pt.pdf} 
    \caption{Feynman diagrams relevant to the calculation in Sec.~\ref{sec:3pt}. The small dot represents the minimal coupling between the complex scalar $\chi$ and the photon, while the large dot represents the dimension-5 $\phi F^2$ interaction.}
    \label{fig:3ptApp}
\end{figure}

The relevant Feynman diagrams for this calculation are given in Fig.~\ref{fig:3ptApp}; the 
the two-particle cut in the $s$-channel receives contributions only from the first diagram.
The sum of these diagrams are free of IR divergences since there is no tree-level amplitude $A_{\phi\chi\to\chi\phi}$ with two $\phi F^{2}$ insertions~\cite{Bern:2020ikv,Catani:1996vz}.
The s-channel cut 
evaluates to
\begin{align}
    \frac{i}{\pi} \textrm{Disc}_s\, \Aoneloop_{\phi\chi\to\chi\phi} &= 
    -\frac{1}{\pi} 
    \frac{c_5^{2}\,g^2}{2}
    \int\text{dLIPS}  \left(\frac{[\ell_1 \,p_2\, \ell_2 \,\ell_1]}{p_1\cdot \ell_1}\right)\left(\frac{\langle\ell_1 \,\ell_2 \, p_3 \,\ell_1\rangle}{p_4\cdot \ell_1}\right)
    +
    (A^+\leftrightarrow A^-)\,,
    \label{S39}
\end{align}
where we use spinor-helicity notation,
$g$ is the coupling of the charged scalar to the photon, and $\text{dLIPS}$ indicates the Lorentz-invariant phase-space integral of $\ell_1$ and $\ell_2$. 
In \eqref{S39}, we only wrote one choice of photon helicities explicitly; the case with swapped helicities ultimately leads to the same contribution as the first term. 
Importantly, the integrand is manifestly positive in the forward limit,
$|[\ell_1 p_2 \ell_2 \ell_1]/(p_1\cdot \ell_1)|^2\geq0$.
However, in contrast to the cases considered in the main text, now the $s$-channel discontinuity 
receives contributions from triangle and box scalar integrals as well,
\begin{align}
    \textrm{Disc}_{s}\Aoneloop_{\phi\chi\to\chi\phi} &= g^{2}c_{5}^{2}\,\textrm{Disc}_{s}\!\lsb -4sI_{\circ}^{(s)} - 4s^{2}I_{\triangle}^{(s)} + 2s^{2}tI_{\square}^{(st)} \rsb.
\end{align}
where the bubble integral, the triangle integral and the box integral, defined as
\begin{align}
    I_{\circ}^{(s)} &=
        \int\frac{d^{d}k}{\lrb 2\pi\rrb^{d}}\frac{1}{k^{2}\lrb k + p_{1} + p_{2}\rrb^{2}}\,,
    \hspace{1.6cm}
    I_{\triangle}^{(s)}=
    \int\frac{d^{d}k}{\lrb 2\pi\rrb^{d}}\frac{1}{k^{2}\lrb k + p_{2}\rrb^{2}\lrb k + p_{1} + p_{2}\rrb^{2}}\,,
    \nn\\
    I_{\square}^{(st)} &=
    \int\frac{d^{d}k}{\lrb 2\pi\rrb^{d}}\frac{1}{k^{2}\lrb k + p_{2}\rrb^{2}\lrb k + p_{1} + p_{2}\rrb^{2}\lrb k - p_{3}\rrb^{2}}\,,
\end{align}
respectively, all contribute to the $s$-channel cut. 
Even though the full $s$-channel cut is necessarily positive in the forward limit (albeit divergent), only the bubble integral has a UV divergence.
Crucially, the bubble  coefficient is negative, showing that a positive integrand of the phase space integral does not necessarily lead to sign-definite contributions to the UV divergence when three-point interactions are present, even if the amplitude is IR-finite.

For completeness, we include the full RG of $c_6$ induced by two insertions of $c_5$ :
\begin{equation}
    \dot{c_{6}} = 24\,g^{2}c_{5}^2\,,
\end{equation}
which is obtained by combining the contributions from all channels.

\subsection{Comparison with Dispersion Relations}

The models with positive $\beta$ functions, and thus the running toward negative values in the IR, may seem to be in tension with the positivity from dispersion relations. We will analyze the dispersion relations at loop level and show that they are in full agreement with the violation of tree-level positivity bounds.

\begin{figure}[t]
  \centering
    \includegraphics[trim={0 25cm 52cm 0},clip,scale=0.5]{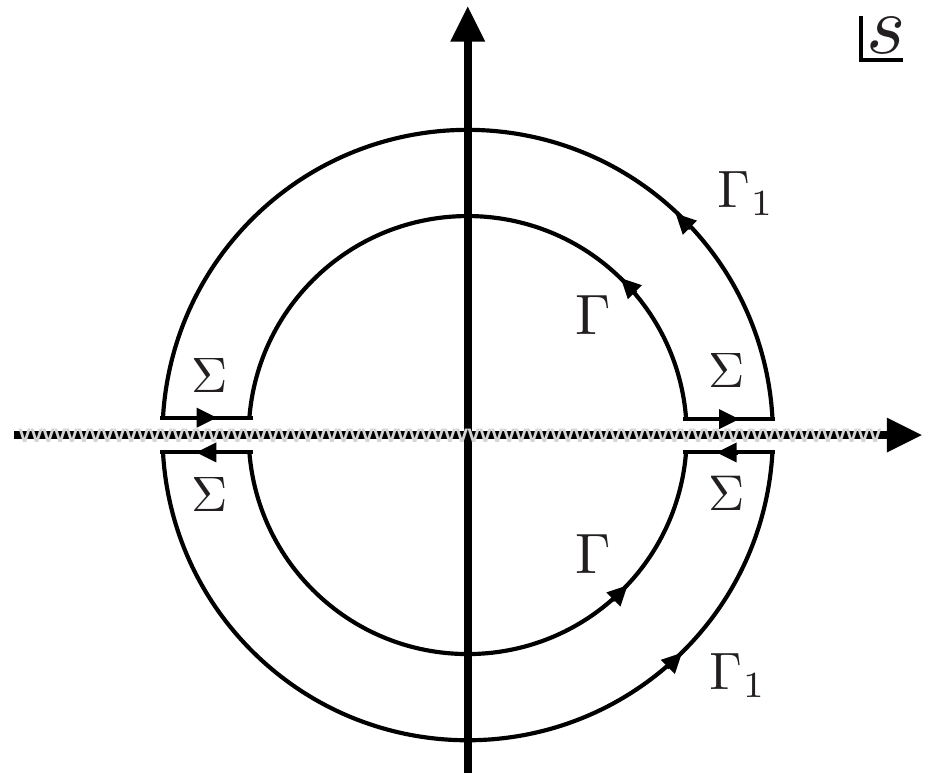}
    
  \caption{Contours in the complex $s$ plane for the dispersion relations. We consider energy within the EFT cutoff $\Lambda$. The grey zigzag lines along the real axis are the branch cuts from $s$ and $u$ channels. The inner and outer circles with radius $r$ and $r_1$ are contours $\Gamma$ and $\Gamma_1$. The contour $\Sigma$ gives the integration along the discontinuity in both positive and negative $s$.}
\label{fig:contours}
\end{figure}

Consider the contours shown in Fig.~\ref{fig:contours} which lie within the EFT regime. We define the contour integral over the arc $\Gamma$ with radius $r$ as
\begin{align}
    \textrm{arc}(r) = \frac{1}{2\pi i}\int_{\Gamma}\frac{ds}{s^{3}}A\left(s,t\right).
\end{align}
By Cauchy's theorem, we have
\begin{align}
    \textrm{arc}(r)=\textrm{arc}(r_1)+\frac{1}{2\pi i}\int_{\Sigma}\frac{ds}{s^{3}}\,\text{Disc} A\!\left(s,t\right),
    \label{eq:Cauchy}
\end{align}
where $\textrm{arc}(r_1)$ is evaluated with the arc $\Gamma_1$ with radius $r_1 > r$. Assuming that the amplitude is symmetric under $s\leftrightarrow u$ and considering the forward limit with $t=0$, we can split the integral over $\Sigma$ as
\begin{align}
    \frac{1}{2\pi i}\int_{\Sigma}\frac{ds}{s^{3}}\,\text{Disc} A\!\left(s,0\right)
    &= \frac{1}{2\pi i}\int^{r_1}_{r_0}\frac{ds}{s^{3}}\,(A(s+i0,0)-A(s-i0,0))
    +\frac{1}{2\pi i}\int^{-r_0}_{-r_1}\frac{ds'}{s^{'3}}\,(A(s'+i0,0)-A(s'-i0,0)) \nonumber \\
    &= \frac{1}{\pi i}\int^{r_1}_{r_0}\frac{ds}{s^{3}}\,(A(s+i0,0)-A(s-i0,0)) \ge 0\,,
\end{align}
where in the second equality, we use the crossing symmetry between $s$ and $u$. The last inequality follows from the optical theorem. We can consider $r_1$ to be infinitesimally larger than $r$. Combining $\int_{\Sigma}\frac{ds}{s^{3}}\,\text{Disc} A\!\left(s,t\right) \ge 0$ and Cauchy's theorem \eqref{eq:Cauchy} yields
\begin{align}
    \frac{d}{dr} \textrm{arc}(r) \le 0\,.
    \label{eq:contour_athm}
\end{align}
This is a constraint derived from unitarity, analyticity and crossing symmetry in the IR regime, but not assuming any other details about the underlying EFT. For instance, the contours do not extend to infinity, so the polynomial boundedness of the amplitude is not needed. This inequality \eqref{eq:contour_athm} can be viewed as the avatar of the \sug{RG positivity} in terms of the dispersion relations.

Let us now consider the amplitude $A\!\left(s,t\right)$ up to one-loop level and dimension eight in an EFT with massless particles
\begin{align}
    A\left(s,t\right)&=\left[-\tilde{\lambda} \nonumber-\frac{\beta_{\tilde{\lambda}}}{6}\left(
    \ln\!\left(\frac{-s}{\mu^2}\right)
    +\ln\!\left(\frac{-u}{\mu^2}\right)
    +\ln\!\left(\frac{-t}{\mu^2}\right)
    +6\right) \right]\nonumber\\
    &+\left[\frac{2\,\tilde{c}_8}{\Lambda^4}(s^2+u^2+t^2)+\frac{\beta_8}{\Lambda^4} \left(
    s^2 \ln\!\left(\frac{-s}{\mu^2}\right)
    +u^2 \ln\!\left(\frac{-u}{\mu^2}\right)
    +t^2 \ln\!\left(\frac{-t}{\mu^2}\right)
    +2\left(s^2+t^2+u^2\right)
    \right)\right] +\dots,
    \label{eq:1loop_amp}
\end{align}
where $\beta_{\tilde{\lambda}}=d\tilde{\lambda}/d\ln \mu$ and $\beta_8=d\tilde{c}_8/d\ln \mu$ are the $\beta$ functions and dots denote higher order terms in either loop and EFT expansion.
We also include the finite terms from the loop integrals using the $\overline{\textrm{MS}}$ subtraction scheme.
See similar analyses including dimension-6 terms in~\cite{Chala:2021wpj,Li:2022aby,Ye:2024rzr,Chala:2023jyx}.
This form of amplitude is very general, since it only assumes crossing symmetry in $s,t,u$ and negligible dimension-6 terms.
One can map \eqref{eq:1loop_amp} to the models in Sec.~\ref{sec:negative_model1} and \ref{sec:negative_model2}, when all external particles are identical under crossing and all dimension-6 couplings are set to zero.
For instance, $(\tilde{\lambda},\tilde{c}_8)=(\lambda,c_8)$ and $(\beta_{\tilde{\lambda}},\beta_8) = (3\lambda^2/(16\pi^2), 5\lambda c_8/(24\pi^2))$ when one considers the EFT in~\eqref{eq:single_phi_EFT}.

By evaluating the contour integrals directly and then taking forward limit, we find
\begin{align}
    \textrm{arc}(r)  &= \frac{\beta_{\tilde{\lambda}}}{6} \frac{1}{r^2}
    +\frac{4}{\Lambda^4}\left( \tilde{c}_8 +\beta_8 
    +\frac{\beta_8}{2} \ln\!\left(\frac{r}{\mu^2}
    \right) \right) + \mathcal{O}\!\left(\frac{r}{\Lambda^{6}}  \right).
    \label{eq:arc_value}
\end{align}
For later convenience, we include an error term in the above from the running of the omitted $\tilde{c}_{10}$. The ratio $r/\Lambda^{6}$ is fixed by dimensional analysis.
The coefficient of $1/\Lambda^4$ is invariant under RG, as it should be since the arc is defined from the amplitude.
Crucially, the running at dimension four is not projected out in $\textrm{arc}(r)$ because it is associated with the non-analytic logarithms.
Taking the derivative of~\eqref{eq:arc_value} leads to the constraint \eqref{eq:contour_athm}
\begin{align}
\frac{d}{dr}\text{arc}(r)=-\frac{\beta_{\tilde{\lambda}}}{3\,r^3}+\frac{2}{\Lambda^4}\frac{\beta_8}{r}
+\mathcal{O}\!\left(\frac{1}{\Lambda^{6}}\right)
\leq0\, ,
\end{align}
which implies
\begin{align}
    \beta_{\tilde{\lambda}}\geq\frac{6\,r^2}{\Lambda^4}\beta_8 +\mathcal{O}\!\left(\frac{r^{3}}{\Lambda^{6}}\right).
    \label{eq:contour_athm_explicit}
\end{align}
To obtain a robust bound without contamination from higher dimensional terms, we should only consider $r \ll \Lambda$. 
If we turn off all the dimension-4 interactions, 
$\beta_{\tilde{\lambda}}=0$, then $\beta_8$ must be negative.
This agrees with our prediction from the theorem derived in the main text,
since $\beta_8$ can only receive contributions from pairs of dimension-6 operators in this case, and the strong-coupling condition in the main text is automatically satisfied.
(When dimension-4 couplings are absent, adding dimension-6 terms does not modify the formal expression of \eqref{eq:contour_athm_explicit}, but only changes the explicit values of $\beta_8$.)
If the running of $\tilde{\lambda}$ is present with positive $\beta_{\tilde{\lambda}}$, then this inequality is satisfied for either sign of $\beta_8$ when $r \ll \Lambda^2$. Using the explicit values in Sec.~\ref{sec:negative_model1} for \eqref{eq:contour_athm_explicit}, we find
\begin{align}
    \lambda \ge \frac{20\,r^2}{3\Lambda^4} c_8 = \frac{r^2}{36\,\Lambda^4}  \frac{g^2}{16\pi^2}\,,
\end{align}
which is satisfied when we consider $g \ll \lambda$. We can see that the positive sign of the $\beta$ function, and thus the running toward smaller or even negative values of $c_8$ in the IR, is in full agreement with the dispersion relations.


\vskip .3 cm

\end{document}